\documentclass[aps,prd,twocolumn,floatfix,nofootinbib,10pt]{revtex4-1}
\usepackage{amsmath,amssymb,multirow,bm,longtable}
\usepackage{graphics,hyperref,subfigure}
\hypersetup{pdfnewwindow=true,
colorlinks=true,linkcolor=blue,
citecolor=blue,filecolor=blue,urlcolor=blue
}
\newcommand{\ba}{\begin{eqnarray}}
\newcommand{\ea}{\end{eqnarray}}
\newcommand{\icn}{Instituto de Ciencias Nucleares, Universidad Nacional Aut\'onoma de M\'exico, Ciudad de M\'exico 04510, Mexico}
\begin{document}
\title{Hidden charm pentaquarks: mass spectrum, magnetic moments, and photocouplings}
\author{Emmanuel Ortiz-Pacheco}
\email{emmanuelo@ciencias.unam.mx}
\affiliation{\icn}
\author{Roelof Bijker}
\email{bijker@nucleares.unam.mx}
\affiliation{\icn}
\author{C\'esar Fern\'andez-Ram{\'{\i}}rez}
\email{cesar.fernandez@nucleares.unam.mx}
\affiliation{\icn}
\begin{abstract}
We develop an extension of the usual 
three flavor quark model to four flavors 
($u$, $d$, $s$ and $c$), 
and discuss the classification of pentaquark 
states with hidden charm ($qqqc\bar{c}$).
We fit our model to the known baryon spectrum and
we predict the double and triple charm baryons,
finding good agreement with the most recent lattice QCD
calculations.
We compute the the ground state of
hidden charm pentaquarks and
their associated magnetic moments and 
electromagnetic couplings, 
of interest to pentaquark photoproduction experiments.
\end{abstract}
\maketitle
\section{Introduction}
\label{sec:introduction}
The discovery and study of
hadron states with quantum numbers 
that go beyond the three-quark picture
is one of the holy grails of hadron spectroscopy because
of the insight that these states 
can provide on confinement, gluonic degrees
of freedom and, in general, on strong QCD in the GeV
scale~\cite{Esposito:2016noz,Olsen:2017bmm,Karliner:2017qhf,Guo:2017jvc}.
If we focus on the baryon sector, the current prime interest is the hidden-charm pentaquark candidates recently found by the LHCb collaboration in the $\Lambda_b^0 \to J/\psi \, p\, K^-$
decay~\cite{Jurik:2016bdm,Aaij:2015tga,Aaij:2016phn}.
The discussion on the nature of the pentaquark candidates has triggered different plausible interpretations of this signals, {\it e.g.}, kinematical effects~\cite{Guo:2015umn}, molecular
states~\cite{Meissner:2015mza,Chen:2015moa,Roca:2015dva,He:2015cea,Lu:2016nnt}, and compact
pentaquarks~\cite{Yuan:2012wz,Maiani:2015vwa,Lebed:2015tna,Mironov:2015ica,Santopinto:2016pkp}.
They have also been studied using diquark-diquark-antiquark interpolating currents combined with QCD sum rules~\cite{Wang:2015epa}.
There are several proposals to confirm their existence using near threshold $J/\psi$
photoproduction~\cite{Wang:2015jsa,Kubarovsky:2015aaa,Karliner:2015voa,Wang:2016dzu,Blin:2016dlf,Fernandez-Ramirez:2017gzc,Blin:2018dkm,HallApc,HallBpc,Meziani:2016lhg,HallDpc}, although such a measurement depends on the size of the electromagnetic coupling of the pentaquarks and
the $J/\psi p$ branching ratio.
Besides confirmation of the signals in a different channel,
to establish them as undisputed pentaquarks
it is needed to find the other members of the multiplet.
A crucial step in this task is to establish a complete classification scheme of the ground state hidden-charm pentaquark configurations of the type $qqqc\bar{c}$ where $q$ denotes the light-quark flavors, $u$, $d$ and $s$. Here we develop a quark model based on four different flavors. Such a scheme provides a complete classification even though if it is broken by the large difference in mass between the charm and the light quarks.

The aim of this paper is threefold. First we discuss an extension of the quark model based on $SU(4)$ flavor symmetry. We develop a mass formula whose coefficients (and their uncertainties) are determined from the mass spectrum of the known $qqq$ and $qqc$ baryons. The same mass formula is applied to calculate the masses of the double and triple charm baryons, $qcc$ and $ccc$. Secondly, we provide a complete classification of ground state hidden-charm pentaquark $qqqc\bar{c}$ states, and show that these belong to either an octet or a decuplet configuration. Finally, we calculate the mass spectrum, magnetic moments, and photocouplings of these pentaquark states which are needed for pentaquark photoproduction studies.

The paper is organized as follows. 
Section~\ref{sec:SU4} reviews the 
extension of the quark model to four flavors.
In Sec.~\ref{sec:3qbaryons} 
we derive the four-flavor mass formula and we 
use the three and four star baryons in the
PDG~\cite{Tanabashi:2018oca} 
to determine the parameters and their uncertainties.
We also provide predictions of double 
and triple charm states which are
benchmarked to the recent LHCb measurement 
of $\Xi_{cc}$ and Lattice QCD (LQCD)
computations.
The uncertainties in the parameters 
of the mass formula are fully carried
to the predictions.
In Sec.~\ref{sec:pentaquarks} we derive the group theory
for hidden-charm pentaquarks and compute
the ground states masses, magnetic moments
and photocouplings.
In Sec.~\ref{sec:conclusions} 
we summarize our main conclusions.
The technical details on the wave functions 
are provided in the Appendices.

\section{Four-flavor SU(4) quark model}
\label{sec:SU4}
In this section, we review the four-flavor $SU(4)$ quark model which is based on the flavors $u$, $d$, $s$ and $c$ \cite{Bjorken:1964gz,Glashow:1970gm,Gaillard:1974mw,DeRujula:1975qlm}. 
It is important to stress that the $SU(4)$ flavor symmetry provides a complete classification scheme for baryons and mesons. The breaking of the $SU(4)$  symmetry is discussed in the next section. 

The spin-flavor states of the $udsc$ quark model can be decomposed into their flavor and spin parts as 
\ba
\left| \begin{array}{cccccc}
& SU_{\rm sf}(8) &\supset& SU_{\rm f}(4) &\otimes& SU_{\rm s}(2) \\
& [f] &,& [g] &,& S 
\end{array} \right> ~,
\ea
followed by a reduction to the three-flavor $uds$ states 
\ba
\left| \begin{array}{cccccc}
& SU_{\rm f}(4) &\supset& SU_{\rm f}(3) &\otimes& U_{\rm Z}(1) \\
& [g] &,& [h] &,& Z 
\end{array} \right> ~,
\ea
and finally to their isospin and hypercharge contents  
\ba
\left| \begin{array}{cccccc}
& SU_{\rm f}(3) &\supset& SU_{\rm I}(2) &\otimes& U_{\rm Y}(1) \\
& [h] &,& I &,& Y
\end{array} \right> ~.
\ea

We make use of the Young tableau technique to construct the allowed representations of multiquark systems, denoting with a box the fundamental representation of $SU(n)$, where $n=2$ for the spin, $n=3$ for the color, $n=3$ ($4$) for the case of three (four) flavors, and $n=6$ ($8$) for the corresponding combined spin-flavor degrees of freedom. 
The quarks transform as the fundamental representation $[1]$ under $SU(n)$, whereas the antiquarks transform as the conjugate representation $[1^{n-1}]$ under $SU(n)$. 

The spin-flavor and flavor states of the four-flavor $udsc$ quark model are labeled by the Young tableaux $[f]$ and $[g]$, respectively. The quarks are associated with the fundamental representations $[f] = [1]$ and $[g] = [1]$, and the antiquarks with their conjugates $[f]=[1111111]\equiv[1^7]$ and $[g]=[111]$, respectively. Similarly, the flavor states of the three-flavor $uds$ quark model are labeled by $[h]$ with 
$[h] = [1]$ for the quark triplet ($u$, $d$, $s$) and $[h] = [0]$ for the singlet ($c$). The antiquarks are associated with the conjugate representations, $[h] = [11]$ for the antitriplet ($\bar{u}$, $\bar{d}$, $\bar{s}$), and $[h] = [111]$ for the singlet ($\bar{c}$). The hypercharges $Y$ and $Z$ are related to the baryon number $B$, the strangeness ${\cal S}$ and charm $C$ by~\cite{Tanabashi:2018oca}
\begin{subequations}
\ba
Y &=& B+{\cal S}-\frac{C}{3} ~,\\
Z &=& \frac{3}{4}B-C ~,
\ea
\end{subequations}
and the electric charge $Q$ is given by the generalized Gell-Mann-Nishijima relation 
\ba
Q &=& I_3+\frac{B+{\cal S}+C}{2} ~.
\ea
Finally, the quarks have spin and parity $S^{\cal P}=1/2^+$ and the antiquarks $1/2^-$. 

In summary, the flavor wave functions of the quarks are characterized by the quantum numbers $\left| [g],[h],I,I_3,Y,Z \right>$ with
\begin{subequations}
\ba 
\left| \phi(u) \right> &=& \left| [1],[1],\frac{1}{2},\frac{1}{2},\frac{1}{3},\frac{1}{4} \right>, \\
\left| \phi(d) \right> &=& \left| [1],[1],\frac{1}{2},-\frac{1}{2},\frac{1}{3},\frac{1}{4} \right>, \\
\left| \phi(s) \right> &=& \left| [1],[1],0,0,-\frac{2}{3},\frac{1}{4} \right>, \\
\left| \phi(c) \right> &=& \left| [1],[0],0,0,0,-\frac{3}{4} \right> ~. 
\ea \label{quarkwf}
\end{subequations}
With the phase convention of Baird and Biedenharn \cite{Baird:1963wv,Haacke:1975rt}, 
the antiquarks are associated with the flavor wave functions
\begin{subequations}
\ba 
\left| \phi(\bar{u}) \right> &=& + \left| [111],[11],\frac{1}{2},-\frac{1}{2},-\frac{1}{3},-\frac{1}{4} \right>,\\
\left| \phi(\bar{d}) \right> &=& - \left| [111],[11],\frac{1}{2}, \frac{1}{2},-\frac{1}{3},-\frac{1}{4} \right> ,\\
\left| \phi(\bar{s}) \right> &=& + \left| [111],[11],0,0,\frac{2}{3},-\frac{1}{4} \right>,\\
\left| \phi(\bar{c}) \right> &=& - \left| [111],[111],0,0,0,\frac{3}{4} \right>.
\ea \label{antiquarkwf}
\end{subequations}
Eqs.~(\ref{quarkwf}) and (\ref{antiquarkwf}) show that the $c$ quark (and $\bar{c}$ antiquark) is distinguished from the light quarks ($u$, $d$, $s$) by the value of the hypercharge $Z$, in the same way as the $s$ quark (and $\bar{s}$ antiquark) is distinguished from the $u$ and $d$ quarks by the hypercharge $Y$.

\section{Three-quark baryons}
\label{sec:3qbaryons}
The spin-flavor states of multiquark systems can be obtained by taking the products of the representations 
of the quarks and/or antiquarks. First, we discuss the case of $qqq$ baryons. There are three possible 
spin-flavor configurations
\begin{equation}
[1]_8 \;\otimes\; [1]_8 \;\otimes\; [1]_8 =
[3]_{120} \;\oplus\; 2 \, [21]_{168} \;\oplus\; [111]_{56},
\label{qqqsu8}
\end{equation}
a symmetric one $[f]=[3]$ with dimension $120$, a mixed-symmetric one $[21]$ with dimension $168$ and 
an antisymmetric one $[111]$ with dimension $56$. 
 
The total baryon wave function which consists of the product of an orbital, flavor, spin and color part 
\ba
\psi = \psi^{\rm o} \phi^{\rm f} \chi^{\rm s} \psi^{\rm c},
\ea
has to be antisymmetric. Since the color part is antisymmetric, the orbital and spin-flavor parts have to have the same symmetry (either symmetric, mixed symmetry or antisymmetric). In the absence of orbital excitations, the spin-flavor part has to be symmetric $[3]_{120}$. The flavor and spin content of the 
symmetric spin-flavor configuration is given in Table~\ref{sfqqq}. There are two flavor multiplets, $[3]$ and $[21]$, both with the same dimension. The former consists of states with spin and parity $S^{\cal P}=3/2^+$ and the latter with $1/2^+$.  

\begin{table}
\centering
\caption[]{Spin-flavor classification of $qqq$ states.}
\label{sfqqq}
\begin{ruledtabular}
\begin{tabular}{cccccc}
$SU_{\rm sf}(8)$ &$\supset$& $SU_{\rm f}(4)$ &$\otimes$& $SU_{\rm s}(2)$ & \\
\noalign{\smallskip}
$[f]$ &$\supset$& $[g]$ &$\otimes$& $[g']$ & $S=\frac{g'_1-g'_2}{2}$ \\
\noalign{\smallskip}
\hline
\noalign{\smallskip}
$[3]_{120}$ && $[3]_{20}$  &$\otimes$& $[3]_4$  & $\frac{3}{2}$ \\
\noalign{\smallskip}
            && $[21]_{20}$ &$\otimes$& $[21]_2$ & $\frac{1}{2}$ \\
\end{tabular}
\end{ruledtabular}
\end{table}

\begin{figure}
\centering
\rotatebox{0}{\scalebox{0.8}[0.8]{\includegraphics{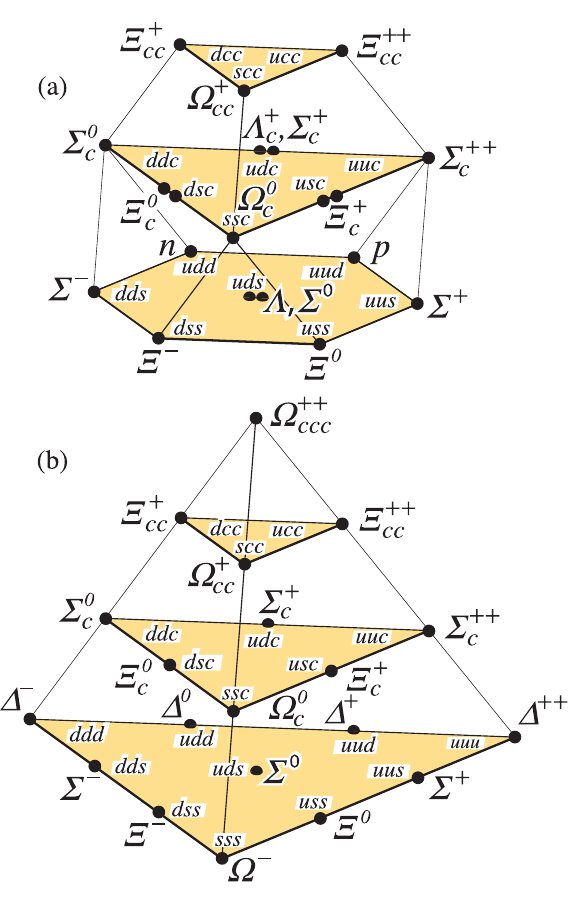}}} 
\caption[]{$SU(4)$ multiplets of $qqq$ baryons made of $u$, $d$, $s$, and $c$ quarks: (a) the $[21]_{20}$ multiplet with $J^P=1/2^+$ containing the baryon octet, and (b) the $[21]_{20}$ multiplet with $J^P=3/2^+$ containing the baryon decuplet (taken from \cite{Tanabashi:2018oca}).} 
\label{baryons}
\end{figure}

The decomposition of four into three flavors $SU_{\rm f}(4) \supset SU_{\rm f}(3) \otimes U_{\rm Z}(1)$ is shown in Table~\ref{fqqq}. The symmetric 20-plet splits into a $uds$ baryon decuplet, a sextet with one charm quark, a triplet with two charm quarks and a singlet consisting of three charm quarks. The 20-plet with mixed symmetry splits into a $uds$ octet, a sextet and an anti-triplet with one charm quark, 
and a triplet with two charm quarks (see Fig.~\ref{baryons}). 

\begin{table}
\centering
\caption[]{$SU_{\rm f}(4) \supset SU_{\rm f}(3) \otimes U_{\rm Z}(1)$ flavor classification of three-quark 
states (here $q$ refers to the light flavors: $u$, $d$, $s$).}
\label{fqqq}
\begin{ruledtabular}
\begin{tabular}{ccccccccc}
$SU_{\rm f}(4)$ &$\supset$& $SU_{\rm f}(3)$ && && && \\
\noalign{\smallskip}
$[g]$ &$\supset$& $[h]$ && && && \\
\noalign{\smallskip}
\hline
\noalign{\smallskip}
$[3]_{20}$   &$\supset$& $[3]_{10}$ &$\oplus$& $[2]_{6}$ &$\oplus$& $[1]_{3}$ &$\oplus$& $[0]_{1}$ \\
\noalign{\smallskip}
$[21]_{20}$  &$\supset$& $[21]_{8}$ &$\oplus$& $[2]_{6} \oplus [11]_{3}$ &$\oplus$& $[1]_{3}$ && \\
\noalign{\smallskip}
\hline
\noalign{\smallskip}
&& $Z=\frac{3}{4}$ && $Z=-\frac{1}{4}$ && $Z=-\frac{5}{4}$ && $Z=-\frac{9}{4}$ \\ 
&& $qqq$ && $qqc$ && $qcc$ && $ccc$
\end{tabular}
\end{ruledtabular}
\end{table}
%

In the literature it is customary to label the spin-flavor and flavor multiplets by their dimension, rather than by their Young tableaux. Table~\ref{sfqqq} shows that for the four-flavor quark model there are two $SU(4)$ flavor multiplets, $[3]_{20}$ and $[21]_{20}$, with the same dimension. 
For this reason, we prefer to use the labeling by Young tableaux. 

In summary, the baryon wave functions are labeled by the quantum numbers
\begin{equation}
\left| [f],(\alpha,\beta,\gamma),(\lambda,\mu),I,Y,Z,L^{\pi},S^{\cal P};J^P \right>.
\label{baryonwf}
\end{equation}
Here $[f]$ denotes the spin-flavor part. For $qqq$ baryons it is either symmetric $[f]=[3]$, mixed symmetry $[21]$ or antisymmetric $[111]$. The orbital part is represented by the orbital angular momentum and parity $L^{\pi}$. Note that the remaining quantum numbers to specify the orbital wave functions uniquely have been suppressed. The flavor part is denoted by the $SU(4)$ multiplets 
\begin{equation}
(\alpha,\beta,\gamma)=(g_1-g_2,g_2-g_3,g_3-g_4) ,
\label{abc}
\end{equation}
the $SU(3)$ multiplets 
\begin{equation}
(\lambda,\mu)=(h_1-h_2,h_2-h_3) , 
\label{lm}
\end{equation}
the isospin $I$, and the hypercharges $Y$ and $Z$. Finally, the total angular momentum is given by the sum of the orbital and the spin parts $\vec{J}=\vec{L}+\vec{S}$. The total parity is given by the product of the parity of the orbital motion and the intrinsic parity of the quarks, $P=\pi {\cal P}$. 

As an example, the nucleon wave function is given by 
\begin{equation}
\left| N(939) \right> = \left| [3],(1,1,0),(1,1),\frac{1}{2},1,\frac{3}{4},0^+,\frac{1}{2}^+;\frac{1}{2}^+ \right>.
\end{equation}
Here $(\alpha,\beta,\gamma)=(1,1,0)$ denotes the $SU(4)$ mixed symmetry 20-plet, and $(\lambda,\mu)=(1,1)$ the $SU(3)$ flavor octet. 

\subsection{Mass formula}
\label{sec:massformula}
In order to study the general structure of the mass spectrum of baryons, we consider a simple schematic model in which the mass operator is given by~\cite{Bijker:1994yr,Bijker:2000gq,Bijker:2003pm}
\ba
M^2 &=& \left( \sum_i m_i \right)^2 + M_{\rm sf}^2 + M_{\rm orb}^2.
\label{msq}
\ea
For the spin-flavor part of the mass operator $M_{\rm sf}^2$ 
we use a generalization of the G\"ursey-Radicati form introduced 
in a study of baryon resonances~\cite{Bijker:1994yr,Bijker:2000gq} 
and pentaquarks~\cite{Bijker:2003pm} to the four-flavor case
\begin{equation}
\begin{split}
M_{\rm sf}^2 = & 
   \, \,\, a \left[ C_{2SU_{\rm sf}(8)}-\frac{231}{16} \right]\\
+ &\, b_3 \left[ C_{2SU_{\rm f}(3)}-3 \right]
+     b_4 \left[ C_{2SU_{\rm f}(4)}-\frac{39}{8} \right] \\
+ &\, c_S \left[ S(S+1)-\frac{3}{4} \right] 
+     c_I \left[ I(I+1)-\frac{3}{4} \right] \\
+ &\, y_1 \left[ Y-1 \right] + y_2 \left[ Y^2-1 \right] \\
+ &\, z_1 \left[ Z-\frac{3}{4} \right]
+     z_2 \left[ Z^2-\frac{9}{16} \right].
\end{split}\label{msqsf}
\end{equation}
The first three terms represent the quadratic Casimir operators of the $SU_{\rm sf}(8)$ spin-flavor and the $SU_{\rm f}(4)$ and $SU_{\rm f}(3)$ flavor groups. 
For the definition of the Casimir operators we have followed the same convention as in~\cite{Stancu:1997dq,Helminen:2000jb} in which the eigenvalue of the quadratic Casimir operator of $SU(n)$ is given by 
\begin{equation}
C_{2SU(n)} = \frac{1}{2} \left[ \sum_{i=1}^n f_i(f_i+n+1-2i) 
- \frac{1}{n} \left(\sum_{i=1}^n f_i\right)^2 \right] .
\label{cas}
\end{equation}

Since in the next section we use the labels $(\lambda,\mu)$ and $(\alpha,\beta,\gamma)$ rather than $[h]$ and $[g]$ for the three- and four-flavor algebras (see Eqs.~\eqref{abc} and \eqref{lm}), we give here the corresponding expressions for the eigenvalues of the Casimir invariants 
\begin{eqnarray}
C_{2SU_f(3)} &=& \frac{1}{3} \left[ \lambda(\lambda+3) + \mu(\mu+3) + \lambda \mu \right] , \\
C_{2SU_f(4)} &=& \frac{1}{2} \left[ \frac{3}{4} \alpha(\alpha+4) + \beta(\beta+4) + \frac{3}{4} \gamma(\gamma+4) \right. 
\nonumber\\
&& \left. + \, \alpha \beta + \beta \gamma + \frac{1}{2} \alpha \gamma \right] .
\label{cas34}
\end{eqnarray}

The last term, $M_{\rm orb}^2$, describes the contribution 
of the orbital degrees of freedom to the mass formula. 
In the present study we only consider the term
\begin{equation}
M_{\rm orb}^2 = \alpha' \, L, \label{msqorb}
\end{equation}
to model the occurrence of linear trajectories in the Chew-Frautschi 
($M^2$ {\it vs} $J$) plots~\cite{Chew:1962eu,Collins:1977jy}. 
By construction, the mass formula of 
Eqs.~\eqref{msq}, \eqref{msqsf} and \eqref{msqorb} 
is defined in such a way that 
$M_{\rm sf}^2 =0$ for the nucleon,
and, hence, its mass is given by 
\begin{equation}
M^2_N = \left( 2m_u+m_d \right)^2 .
\label{nucleonmass}
\end{equation}

In many studies of multiquark configurations, 
effective spin-flavor hyperfine interactions have been used in the 
constituent quark model (CQM) 
which schematically represents the Goldstone boson exchange
interaction between constituent
quarks~\cite{Stancu:1997dq,Stancu:1991rc,Helminen:2000jb,Carlson:2003pn,Carlson:2003wc,Glozman:2003sy}. 
An analysis of the strange and non-strange $qqq$ baryon resonances 
in the collective stringlike model~\cite{Bijker:1994yr,Bijker:2000gq} 
and the hypercentral CQM~\cite{Giannini:2015zia} also showed evidence for the need 
of such type of interaction terms. If one neglects the radial dependence, 
the matrix elements of these interactions for the general case of $k$ flavors 
depend on the Casimirs of the $SU_{\rm sf}(2k)$ spin-flavor, 
the $SU_{\rm f}(k)$ flavor and 
the $SU_{\rm s}(2)$ spin groups as 
\begin{widetext}
\begin{equation}
\left< \sum_{i<j}^n (\vec{\lambda}_i \cdot \vec{\lambda}_j) 
(\vec{\sigma}_i \cdot \vec{\sigma}_j) \right> \;=\; 
4\, C_{2SU_{\rm sf}(2k)} - 2\, C_{2SU_{\rm f}(k)} 
-\frac{4}{k} S(S+1)-n\frac{3(k^2-1)}{k} ~,
\label{hyperfine}
\end{equation}
\end{widetext}
where $n$ is the number of quarks, $\vec{\sigma}$ denotes the three Pauli spin matrices, and $\vec{\lambda}$ the $k^2-1$ Gell-Mann flavor matrices. 
For two flavors ($k=2$), the three flavor matrices are those of the isospin $\vec{\tau}$, and Eq.~\eqref{hyperfine} reduces to Wigner's $SU(4)$ supermultiplet theory~\cite{Wigner:1936dx,talmi1993simple}. 
For three flavors ($k=3$), the matrix elements of the hyperfine interaction are the same as those given by~\cite{Helminen:2000jb}. The present case of interest is for four flavors ($k=4$). 

The energy splittings within a spin-flavor multiplet into four-flavor multiplets and spin states are very similar to that of the $a$, $b_4$ and $c_S$ terms of the mass formula of Eq.~\eqref{msqsf}. In addition, Eq.~\eqref{msqsf} contains the splittings of the four-flavor $SU_{\rm f}(4)$ multiplets according to the generalized Gell-Mann-Okubo formula (see Eq.~\eqref{GMO4} of Appendix~\ref{app:GMO}) via the $b_3$, $z_1$ and $z_2$ terms, as well as the splittings of the three-flavor $SU_{\rm f}(3)$ multiplets of the usual Gell-Mann-Okubo formula of Eq.~\eqref{GMO3} due the $c_I$, $y_1$ and $y_2$ terms. 

In summary, the form of the mass operator proposed in Eqs.~\eqref{msq}, \eqref{msqsf} and \eqref{msqorb} is motivated by previous studies of the mass spectrum of baryon resonances for three flavors in which it was found that a G\"ursey-Radicati form gives a good description of the baryon spectrum~\cite{Bijker:1994yr,Bijker:2000gq}. Here we generalized this form to the case of four quark flavors. The different terms in the mass formula of Eqs.~\eqref{msq}, \eqref{msqsf} and \eqref{msqorb} correspond to the quark masses ($m_u$, $m_d$, $m_s$ and $m_c$), the spin-flavor hyperfine splittings ($a$, $b_4$ and $c_S$), the Gell-Mann-Okubo mass formula for four flavors ($b_3$, $z_1$ and $z_2$) and for three flavors ($c_I$, $y_1$ and $y_2$), and the trajectory slope ($\alpha'$). 

Even though the strong interaction does not distinguish between quarks of different flavor, the $SU_{\rm f}(4)$ flavor symmetry is broken dynamically by the quark masses and the $z_1$ and $z_2$ terms. The latter are proportional to the hypercharge $Z$ which distinguishes the $c$ quark from the light quarks. 
These terms give an additional contribution to the mass but do not change the flavor wave function since they are proportional to Casimir invariants. In the same way, the $SU(3)$ flavor symmetry is broken dynamically in the Gell-Mann-Okubo mass formula by the quark masses and the term proportional to the hypercharge $Y$. 

\subsection{Determination of the parameters}
\label{sec:fit}
To determine the parameters of the mass formula and their 
uncertainties we fit the Breit-Wigner masses of the three- and 
four-star resonances in the PDG~\cite{Tanabashi:2018oca}, namely the 
resonances in Tables~\ref{tab:groundstates} (ground states 
without radial excitations) and~\ref{tab:excitedstates} 
(excited states).\footnote{We also performed the fits using the 
pole masses instead of the Breit-Wigner masses if they are 
available. The results stayed the same.}
We do not include the $N(1440)$, $N(1710)$, $\Delta(1600)$, 
$\Sigma(1660)$, $\Sigma(1940)$, $\Lambda(1600)$ and 
$\Lambda(1810)$ resonances in our fits because there were 
identified as vibrational excitations in previous studies using 
a similar mass formula for the 
three-flavor case~\cite{Bijker:1994yr,Bijker:2000gq}.
Hence, we consider $52$ states to determine the 14 parameters of the mass formula in Eq.~\eqref{msq};º {\it i.e.} 
the constituent quark masses $m_u$, $m_d$, $m_s$ and $m_c$, 
the couplings
$a$, $b_3$, $b_4$, $c_S$, $c_I$, $y_1$, $y_2$, $z_1$ and $z_2$, 
and the slope $\alpha'$.
The parameters $m_u$, $m_d$, $m_c$ and $z_2$ are determined in the following way:
\begin{itemize}
\item[--] According to Eq.~\eqref{nucleonmass}, the mass of the nucleon only depends on $m_u$ and $m_d$ (from the first term in Eq.~\eqref{msq}, the other terms give a vanishing contribution). In our calculations, the masses of the up and down constituent quarks are taken to be identical and fixed by the 
nucleon mass $m_N$, 
$m_u=m_d \equiv m_n = m_N/3=312.972~\text{MeV}$;
\item[--] We do not include information on double and triple charm states in our fits because these resonances have a one star status in the PDG.
Hence, parameters $z_1$ and $z_2$ cannot be determined independently. Inspection of the quantum numbers in Tables~\ref{tab:groundstates} and~\ref{tab:excitedstates} shows that only a linear combination of $z_1$ and $z_2$ can be determined. In our calculations we set $z_2=0$. The impact and reasonability of this choice will be addressed in Sec.~\ref{sec:23cstates};
\item[--] The precise mass of the constituent charm quark $m_c$ is unknown and varies widely from one model to another~\cite{Bijker:2000gq,Molina:2017iaa}. 
Moreover, we found that its value is strongly correlated with $z_1$. We decided not to fit it to the data but rather fix it within a wide range of values, randomly selecting a value for the mass between $1400$ and $1500$ MeV according to an uniform distribution.
This error is propagated to the fits and predictions through a Monte Carlo and constitutes one of the main sources of uncertainties
in the predictions.
\end{itemize}

\begin{table*}
\caption{Ground baryon states (octet and decuplet) fitted to determine the parameters of the mass formula of Eqs.~\eqref{msq}, \eqref{msqsf} and \eqref{msqorb}. The states are labeled by the quantum numbers in Eqs.~\eqref{baryonwf}, \eqref{abc}, and \eqref{lm}. All states have $L^{\pi}=0^+$. $N_q$ denotes the number of up and down quarks ($q=n$), strange quarks ($q=s$) and charm quarks ($q=c$). 
Experimental masses are taken from~\cite{Tanabashi:2018oca} averaging charged and neutral states when needed. The last column shows the deviation 
between the experimental and theoretical mean values
$D= 100 \left( \bar{M}^{exp}-\bar{M}^{th} \right) \slash \bar{M}^{exp}$.} \label{tab:groundstates}
\begin{ruledtabular}
\begin{tabular}{lcccccrrcccccc}
Name & $M^{exp}$ (MeV) & $[f]$&$(\alpha, \beta, \gamma)$ & $(\lambda, \mu)$ 
& $I$ & $Y$ & $Z$ & $S^{\cal P}=J^P$ & $(N_n,N_s,N_c)$ & $M^{th}$ (MeV)& $D$ (\%)\\
\hline
$N(939)$ & $\phantom{0}938.92 \pm 0.91$ & $[3]$ & $(1,1,0)$ & $(1,1)$ 
& $\frac{1}{2}$ & $1$ & $\frac{3}{4}$ & $\frac{1}{2}^+$ & $(3,\, 0,\, 0)$ 
& $\phantom{0}938.92$ (fixed) & $\phantom{-}0.00$\\
$\Sigma(1193)$ & $1193.2 \pm 4.1$ & $[3]$ & $(1,1,0)$ & $(1,1)$ 
& $1$ & $0$ & $\frac{3}{4}$ & $\frac{1}{2}^+$ & $(2,\: 1,\: 0)$ 
& $1161.7 \pm 3.4$ & $\phantom{-}2.64$\\ 
$\Lambda(1116)$ & $1115.683 \pm 0.006$ & $[3]$ & $(1,1,0)$ & $(1,1)$ 
& $0$ & $0$ & $\frac{3}{4}$ & $\frac{1}{2}^+$ & $(2,\: 1,\: 0)$ 
& $1129.6 \pm 2.3 $ & $-1.25$\\  
$\Xi(1318)$ & $1318.3 \pm 4.8$ & $[3]$ & $(1,1,0)$ & $(1,1)$ 
& $\frac{1}{2}$ & $-1$ & $\frac{3}{4}$ & $\frac{1}{2}^+$ & $(1,\: 2,\: 0 )$ 
& $1321.8 \pm 2.8 $ & $-0.27$\\
$\Sigma_c(2455)$ & $2453.50 \pm 0.57$ & $[3]$ & $(1,1,0)$ & $(2,0)$ 
& $1$ & $\frac{2}{3}$ & $-\frac{1}{4}$ & $\frac{1}{2}^+$ & $(2,\: 0,\: 1)$ 
& $2419.88 \pm 0.96$ & $\phantom{-} 1.37$\\   
$\Lambda_c$ & $2286.46 \pm 0.14$ & $[3]$ & $(1,1,0)$ & $(0,1)$ 
& $0$ & $\frac{2}{3}$ & $-\frac{1}{4}$ & $\frac{1}{2}^+$ & $(2,\: 0,\: 1)$ 
& $ 2357.91 \pm 0.96 $ & $-3.12$\\   
$\Xi_c$ & $2576.8 \pm 1.6$ & $[3]$ & $(1,1,0)$ & $(2,0)$ 
& $\frac{1}{2}$ & $-\frac{1}{3}$ & $-\frac{1}{4}$ & $\frac{1}{2}^+$ & $(1,\: 1,\: 1 )$ 
& $2561.90 \pm  0.94$ & $\phantom{-}0.58 $\\
$\Xi_c'$ & $2469.4 \pm 2.1$ & $[3]$ & $(1,1,0)$ & $(0,1)$ 
& $\frac{1}{2}$ & $-\frac{1}{3}$ & $-\frac{1}{4}$ & $\frac{1}{2}^+$ & $(1,\: 1,\: 1)$ 
& $2518.1 \pm 1.1$ & $-1.97$\\
$\Omega_c$ & $2695.2 \pm 1.7$ & $[3]$ & $(1,1,0)$ & $(2,0)$ 
& $0$ & $-\frac{4}{3}$ & $-\frac{1}{4}$ & $\frac{1}{2}^+$ & $(0,\: 2,\: 1)$ 
& $2703.8 \pm 1.2$ & $-0.32$\\   
$\Delta(1232)$ & $1232.0 \pm 2.0$ & $[3]$ & $(3,0,0)$ & $(3,0)$ 
& $\frac{3}{2}$ & $1$ & $\frac{3}{4}$ & $\frac{3}{2}^+$ & $(3,\: 0,\: 0 )$ 
& $ 1265.1 \pm 3.8$ & $-2.69$\\
$\Sigma^*(1385)$ & $1384.6 \pm 2.3$ & $[3]$ & $(3,0,0)$ & $(3,0)$ 
& $1$ & $0$ & $\frac{3}{4}$ & $\frac{3}{2}^+$ & $(2,\: 1,\: 0)$ 
& $1399.4 \pm 2.3$ & $-1.07$\\   
$ \Xi^*(1530)$ & $1533.4 \pm 2.3$ & $[3]$ & $(3,0,0)$ & $(3,0)$ 
& $\frac{1}{2}$ & $-1$ & $\frac{3}{4}$ & $\frac{3}{2}^+$ & $(1,\: 2,\: 0)$ 
& $1535.0 \pm 1.7$ & $-0.10$\\
$\Omega(1672)$ & $1672.45 \pm 0.29$ & $[3]$ & $(3,0,0)$ & $(3,0)$ 
& $0$ & $-2$ & $\frac{3}{4}$ & $\frac{3}{2}^+$ & $(0,\: 3,\: 0)$ 
& $1671.4 \pm 1.5$ & $\phantom{-} 0.06$\\  
$\Sigma_c^*(2520)$ & $2518.13 \pm 0.55$ & $[3]$ & $(3,0,0)$ & $(2,0)$ 
& $1$ & $\frac{2}{3}$ & $-\frac{1}{4}$ & $\frac{3}{2}^+$ & $(2,\: 0,\: 1)$ 
& $ 2476.02 \pm 0.83$ & $\phantom{-} 1.67$\\ 
$\Xi_c^*(2645)$ & $2645.90 \pm 0.50$ & $[3]$ & $(3,0,0)$ & $(2,0)$ 
& $\frac{1}{2}$ & $-\frac{1}{3}$ & $-\frac{1}{4}$ & $\frac{3}{2}^+$ & $(1,\: 1,\: 1 )$ 
& $2614.99 \pm 0.86$ & $\phantom{-} 1.17$\\
$\Omega_c^*(2770)$ & $2765.9 \pm 2.0$ & $[3]$ & $(3,0,0)$ & $(2,0)$ 
& $0$ & $-\frac{4}{3}$ & $-\frac{1}{4}$ & $\frac{3}{2}^+$ & $(0,\: 2,\: 1)$ 
& $2754.2 \pm 1.2 $ & $\phantom{-} 0.42$
\end{tabular}
\end{ruledtabular}
\end{table*}

The remaining 10 parameters and their uncertainties
are determined by fitting the resonances 
in Tables~\ref{tab:groundstates} and~\ref{tab:excitedstates} 
with the exception of the nucleon (used to fix $m_n$), 
for a total of $N_r=51$ states. 
In doing so, we use the bootstrap 
technique~\cite{recipes,EfroTibs93,Landay:2016cjw}
and proceed as 
follows~\cite{Fernandez-Ramirez:2015fbq,Molina:2017iaa}:

\begin{table*}
\caption{Excited baryon states fitted to determine the parameters of the mass formula of Eqs.~\eqref{msq}, \eqref{msqsf} and \eqref{msqorb}. 
Notation as in Table~\ref{tab:groundstates}.} 
\label{tab:excitedstates}
\begin{ruledtabular}
\begin{tabular}{lcccccrrcccccccc}
Name & $M^{exp}$ (MeV) & $[f]$ & $(\alpha,\beta,\gamma)$ & $(\lambda,\mu)$ 
& $I$ & $Y$ & $Z$ & $L^{\pi}$ & $S^{\cal P}$ & $J^P$ & $(N_n,N_s,N_c)$ & $M^{th}$ (MeV) & $D$ (\%)\\
\hline
$\Lambda(1405)$ & $1405.1 \pm 1.2$ & $[21]$ & $(0,0,1)$ & $(0,0)$
& $0$ & $0$ & $\frac{3}{4}$ & $1^-$ & $\frac{1}{2}^+$ & $\frac{1}{2}^-$ & $(2,\: 1,\: 0)$ 
& $\phantom{0}1600 \pm 10$ & $-13.88\phantom{0}$ \\
$\Lambda(1520)$ & $1519.5 \pm 1.0$ & $[21]$ & $(0,0,1)$ & $(0,0)$
& $0$ & $0$ & $\frac{3}{4}$ & $1^-$ & $\frac{1}{2}^+$ & $\frac{3}{2}^-$ & $(2\: ,1\: ,0)$ 
& $\phantom{0}1600 \pm 10$ & $-5.30$ \\
$\Lambda(2100)$ & $\phantom{0}2100 \pm 10$ & $[21]$ & $(0,0,1)$ & $(0,0)$
& $0$ & $0$ & $\frac{3}{4}$ & $3^-$ & $\frac{1}{2}^+$ & $\frac{7}{2}^-$ & $(2,\: 1,\: 0)$ 
& $\phantom{0}2166 \pm 12$ & $-3.17$ \\
$N(1520)$ & $1515.0 \pm 5.0$ & $[21]$ & $(1,1,0)$ & $(1,1)$ 
& $\frac{1}{2}$ & $1$ & $\frac{3}{4}$ & $1^-$ & $\frac{1}{2}^+$ & $\frac{3}{2}^-$ & $(3,\: 0,\: 0)$ 
& $1557.6 \pm 8.9$ & $-2.81$ \\
$N(1535)$ & $\phantom{0}1535 \pm 10$ & $[21]$ & $(1,1,0)$ & $(1,1)$ 
& $\frac{1}{2}$ & $1$ & $\frac{3}{4}$ & $1^-$ & $\frac{1}{2}^+$ & $\frac{1}{2}^-$ & $(3,\: 0,\: 0)$ 
& $1557.6 \pm 8.9$ & $-1.47$ \\
$N(1650)$ & $\phantom{0} 1658 \pm 12$ & $[21]$ & $(1,1,0)$ & $(1,1)$ 
& $\frac{1}{2}$ & $1$ & $\frac{3}{4}$ & $1^-$ & $\frac{3}{2}^+$ & $\frac{1}{2}^-$ & $(3,\: 0,\: 0)$ 
& $\phantom{0}1666 \pm 10$ & $-0.49$ \\
$N(1675)$ & $1675.0 \pm 5.0$ & $[21]$ & $(1,1,0)$ & $(1,1)$ 
& $\frac{1}{2}$ & $1$ & $\frac{3}{4}$ & $1^-$ & $\frac{3}{2}^+$ & $\frac{5}{2}^-$ & $(3,\: 0,\: 0)$ 
& $\phantom{0}1666 \pm 10$ & $\phantom{-}0.56$\\
$N(1680)$ & $1685.0 \pm 5.0$ & $[3]$ & $(1,1,0)$ & $(1,1)$ 
& $\frac{1}{2}$ & $1$ & $\frac{3}{4}$ & $2^+$ & $\frac{1}{2}^+$ & $\frac{5}{2}^+$ & $(3,\: 0,\: 0)$ 
& $\phantom{0}1736 \pm 12$ & $-3.05$\\
$N(1700)$ & $\phantom{0}1700 \pm 50$ & $[21]$ & $(1,1,0)$ & $(1,1)$
& $\frac{1}{2}$ & $1$ & $\frac{3}{4}$ & $1^-$ & $\frac{3}{2}^+$ & $\frac{3}{2}^-$ & $(3,\: 0,\: 0)$ 
& $\phantom{0}1666 \pm 10$ & $\phantom{-}2.02$ \\
$N(1720)$ & $\phantom{0}1725 \pm 25$ & $[3]$ & $(1,1,0)$ & $(1,1)$
& $\frac{1}{2}$ & $1$ & $\frac{3}{4}$ & $2^+$ & $\frac{1}{2}^+$ & $\frac{3}{2}^+$ & $(3,\: 0,\: 0)$ 
& $\phantom{0}1736 \pm 12$ & $-0.66$ \\
$N(2190)$ & $\phantom{0}2150 \pm 50$ & $[21]$ & $(1,1,0)$ & $(1,1)$
& $\frac{1}{2}$ & $1$ & $\frac{3}{4}$ & $3^-$ & $\frac{1}{2}^+$ & $\frac{7}{2}^-$ & $(3,\: 0,\: 0)$ 
& $\phantom{0}2135 \pm 12$ & $\phantom{-}0.68$ \\
$N(2220)$ & $\phantom{0}2250 \pm 50$ & $[3]$ & $(1,1,0)$ & $(1,1)$
& $\frac{1}{2}$ & $1$ & $\frac{3}{4}$ & $4^+$ & $\frac{1}{2}^+$ & $\frac{9}{2}^+$ & $(3,\: 0,\: 0)$ 
& $\phantom{0}2269 \pm 19 $ & $-0.84$ \\
$N(2250)$ & $\phantom{0}2285 \pm 35$ & $[21]$ & $(1,1,0)$ & $(1,1)$
& $\frac{1}{2}$ & $1$ & $\frac{3}{4}$ & $3^-$ & $\frac{3}{2}^+$ & $\frac{9}{2}^-$ & $(3,\: 0,\: 0)$ 
& $\phantom{0}2215 \pm 12$& $\phantom{-}3.05$ \\
$N(2600)$ & $\phantom{00}2650\pm 100$ & $[21]$ & $(1,1,0)$ & $(1,1)$
& $\frac{1}{2}$ & $1$ & $\frac{3}{4}$ & $5^-$ & $\frac{1}{2}^+$ & $\frac{11}{2}^-$ & $(3,\: 0,\: 0)$ 
& $\phantom{0}2587 \pm  18$ & $\phantom{-}2.38$ \\
$\Lambda(1670)$ & $\phantom{0}1670 \pm 10$ & $[21]$ & $(1,1,0)$ & $(1,1)$
& $0$ & $0$ & $\frac{3}{4}$ & $1^-$ & $\frac{1}{2}^+$ & $\frac{1}{2}^-$ & $(2,\: 1,\: 0)$ 
& $1679.5 \pm 8.2$ & $-0.57$ \\
$\Lambda(1690)$ & $1690.0 \pm 5.0$ & $[21]$ & $(1,1,0)$ & $(1,1)$
& $0$ & $0$ & $\frac{3}{4}$ & $1^-$ & $\frac{1}{2}^+$ & $\frac{3}{2}^-$ & $(2,\: 1,\: 0)$  
& $1679.5 \pm 8.2$ & $\phantom{-}0.62$ \\
$\Lambda(1800)$ & $1785.0 \pm 6.5$ & $[21]$ & $(1,1,0)$ & $(1,1)$
& $0$ & $0$ & $\frac{3}{4}$ & $1^-$ & $\frac{3}{2}^+$ & $\frac{1}{2}^-$ & $(2,\: 1,\: 0)$  
& $1780.1 \pm 9.9$ & $\phantom{-}0.27$ \\
$\Lambda(1820)$ & $1820.0 \pm 5.0$ & $[3]$ & $(1,1,0)$ & $(1,1)$
& $0$ & $0$ & $\frac{3}{4}$ & $2^+$ & $\frac{1}{2}^+$ & $\frac{5}{2}^+$ & $(2,\: 1,\: 0)$
& $\phantom{0}1846 \pm 11$ & $-1.45$ \\
$\Lambda(1830)$ & $\phantom{0}1820 \pm 10$ & $[21]$ & $(1,1,0)$ & $(1,1)$
& $0$ & $0$ & $\frac{3}{4}$ & $1^-$ & $\frac{3}{2}^+$ & $\frac{5}{2}^-$ & $(2,\: 1,\: 0)$
& $1780.1 \pm 9.9$ & $\phantom{-}2.19$ \\
$\Lambda(1890)$ & $\phantom{0}1880 \pm 30$ & $[3]$ & $(1,1,0)$ & $(1,1)$
& $0$ & $0$ & $\frac{3}{4}$ & $2^+$ & $\frac{1}{2}^+$ & $\frac{3}{2}^+$ & $(2,\: 1,\: 0)$
& $\phantom{0}1846 \pm 11$ & $\phantom{-}1.78$ \\  
$\Lambda(2110)$ & $\phantom{0}2115 \pm 25$ & $[21]$ & $(1,1,0)$ & $( 1,1)$
& $0$ & $0$ & $\frac{3}{4}$ & $2^+$ & $\frac{3}{2}^+$ & $\frac{5}{2}^+$ & $(2,\: 1,\: 0)$
& $2058.1 \pm 9.5$ & $\phantom{-}2.69$ \\
$\Lambda(2350)$ & $\phantom{0}2355 \pm 15$ & $[3]$ & $(1,1,0)$ & $(1,1)$
& $0$ & $0$ & $\frac{3}{4}$ & $4^+$ & $\frac{1}{2}^+$ & $\frac{9}{2}^+$ & $(2,\: 1,\: 0)$
& $\phantom{0}2354 \pm 17$ & $\phantom{-}0.03$ \\
$\Xi(1820)$ & $1823.0 \pm 5.0$ & $[21]$ & $(1,1,0)$ & $(1,1)$
& $\frac{1}{2}$ & $-1$ & $\frac{3}{4}$ & $1^-$ & $\frac{1}{2}^+$ & $\frac{3}{2}^-$ & $(1,\: 2,\: 0)$  
& $1814.3 \pm 7.3$ & $\phantom{-}0.47$ \\
$\Sigma(1670)$ & $1677.5 \pm 7.5$ & $[21]$ & $(1,1,0)$ & $(1,1)$
& $1$ & $0$ & $\frac{3}{4}$ & $1^-$ & $\frac{1}{2}^+$ & $\frac{3}{2}^-$ & $(2,\: 1,\: 0)$   
& $1701.2 \pm 7.7$ & $-1.41$ \\
$\Sigma(1750)$ & $\phantom{0}1765 \pm 35$ & $[21]$ & $(1,1,0)$ & $(1,1)$
& $1$ & $0$ & $\frac{3}{4}$ & $1^-$ & $\frac{1}{2}^+$ & $\frac{1}{2}^-$ & $(2,\: 1,\: 0)$ 
& $1701.2 \pm 7.7$ & $\phantom{-}3.61$ \\
$\Sigma(1775)$ & $1775.0 \pm 5.0$ & $[21]$ & $(1,1,0)$ & $(1,1)$
& $1$ & $0$ & $\frac{3}{4}$ & $1^-$ & $\frac{3}{2}^+$ & $\frac{5}{2}^-$ & $(2,\: 1,\: 0)$   
& $1800.6 \pm 9.2$ & $-1.44$ \\
$\Sigma(1915)$ & $\phantom{0}1918 \pm 18$ & $[3]$ & $(1,1,0)$ & $(1,1)$
& $1$ & $0$ & $\frac{3}{4}$ & $2^+$ & $\frac{1}{2}^+$ & $\frac{5}{2}^+$ & $(2,\: 1,\: 0)$ 
& $1800.6 \pm 9.2$ & $\phantom{-}2.67$ \\
$\Delta(1620)$ & $\phantom{0}1630 \pm 30$ & $[21]$ & $(3,0,0)$ & $(3,0)$
& $\frac{3}{2}$ & $1$ & $\frac{3}{4}$ & $1^-$ & $\frac{1}{2}^+$ & $\frac{1}{2}^-$ & $(3,\: 0,\: 0)$ 
& $\phantom{0}1672 \pm 17$ & $-2.60$  \\
$\Delta(1700)$ & $\phantom{0}1710 \pm 40$ & $[21]$ & $(3,0,0)$ & $(3,0)$
& $\frac{3}{2}$ & $1$ & $\frac{3}{4}$ & $1^-$ & $\frac{1}{2}^+$ & $\frac{3}{2}^-$ & $(3,\: 0,\: 0)$  
& $\phantom{0}1672 \pm 17$ & $\phantom{-}2.20$ \\
$\Delta(1905)$ & $\phantom{0}1882 \pm 28$ & $[3]$ & $(3,0,0)$ & $(3,0)$
& $\frac{3}{2}$ & $1$ & $\frac{3}{4}$ & $2^+$ & $\frac{3}{2}^+$ & $\frac{5}{2}^+$ & $(3,\: 0,\: 0)$ 
& $\phantom{0}1932 \pm 11$ & $-2.65$ \\
$\Delta(1910)$ & $\phantom{0}1885 \pm 25$ & $[3]$ & $(3,0,0)$ & $(3,0)$
& $\frac{3}{2}$ & $1$ & $\frac{3}{4}$ & $2^+$ & $\frac{3}{2}^+$ & $\frac{1}{2}^+$ & $(3,\: 0,\: 0)$ 
& $\phantom{0}1932 \pm 11$ & $-2.51$ \\
$\Delta(1920)$ & $\phantom{0}1935 \pm 35$ & $[3]$ & $(3,0,0)$ & $(3,0)$
& $\frac{3}{2}$ & $1$ & $\frac{3}{4}$ & $2^+$ & $\frac{3}{2}^+$ & $\frac{3}{2}^+$ & $(3,\: 0,\: 0)$ 
& $\phantom{0}1932 \pm 11$ & $\phantom{-}0.14$ \\
$\Delta(1930)$ & $\phantom{0}1950 \pm 50$ & $[21]$ & $(3,0,0)$ & $(3,0)$
& $\frac{3}{2}$ & $1$ & $\frac{3}{4}$ & $2^-$ & $\frac{1}{2}^+$ & $\frac{5}{2}^-$ & $(3,\: 0,\: 0)$ 
& $\phantom{0}1966 \pm 16$ & $-0.80$ \\
$\Delta(1950)$ & $\phantom{0}1932 \pm 18$ & $[3]$ & $(3,0,0)$ & $(3,0)$
& $\frac{3}{2}$ & $1$ & $\frac{3}{4}$ & $2^+$ & $\frac{3}{2}^+$ & $\frac{7}{2}^+$ & $(3,\: 0,\: 0)$ 
& $\phantom{0}1932 \pm 11$ & $\phantom{-}0.01$ \\
$\Delta(2420)$ & $\phantom{00}2400 \pm 100$ & $[3]$ & $(3,0,0)$ & $(3,0)$
& $\frac{3}{2}$ & $1$ & $\frac{3}{4}$ & $4^+$ & $\frac{3}{2}^+$ & $\frac{11}{2}^+$ & $(3,\: 0,\: 0)$ 
& $\phantom{0}2422 \pm 17$ & $-0.93$ \\
$\Sigma(2030)$ & $2032.5 \pm 7.5$ & $[3]$ & $(3,0,0)$ & $(3,0)$
& $1$ & $0$ & $\frac{3}{4}$ & $2^+$ & $\frac{3}{2}^+$ & $\frac{7}{2}^+$ & $(2,\: 1,\: 0)$ 
& $2022.8 \pm 9.9$ & $\phantom{-}0.48$ 
\end{tabular}
\end{ruledtabular}
\end{table*}
\begin{enumerate}
\item First, we sample the masses of the resonances ($M^{exp}$), given in Tables~\ref{tab:groundstates} and~\ref{tab:excitedstates}, according to Gaussian distributions whose widths correspond to the experimental uncertainties, obtaining a resampled spectrum;
\item We minimize the weighted squared distance
\begin{equation}
\delta^2 = \sum_{i=1}^{N_r} \omega_i 
\left( M^{exp}_{i} - M^{th}_i \right)^2,  
\nonumber
\end{equation}
where $\omega_i$ is a weight set equal to one for all the states except for the octet and decuplet ground states (Table~\ref{tab:groundstates})
which is set to $\omega_{octet}=\omega_{decuplet}=10$
and the singlet states $\Lambda(1405)$, $\Lambda(1520)$ and $\Lambda(2100)$ which is set to $\omega_\text{singlet}=1/2$.
Regarding the $\Lambda$ singlets, we know that these states are poorly reproduced by quark models unless strong dynamical
effects are included~\cite{Santopinto:2014opa,Faustov:2015eba}.
Hence, because we are interested in the average 
description of the spectrum, we downplay their impact in the fits.
The fits are performed using {\tt MINUIT}~\cite{James:1975dr}
and a total of $N_r=51$ states are fitted with $n_p=10$ parameters, namely: $m_s$, $a$, $b_3$, $b_4$, $c_S$, $c_I$, $y_1$, $y_2$, $z_1$ and $\alpha'$;
\item We repeat this process $10^4$ times providing 
enough statistics to compute the mean value of the parameters and their uncertainties (standard deviation);
\item Next, we can compute the theoretical masses of the resonances and their uncertainties.
In doing so, for each one of the $10^4$ fits we compute the corresponding mass using Eqs.~\eqref{msq}, \eqref{msqsf} and \eqref{msqorb}. 
Then the mass of the resonance is computed as the mean value ($\bar{M}^{th}$) and the uncertainty as the standard deviation.
In this way the uncertainties in the parameters as well as their correlations are exactly propagated to the masses.
In the same way we can make predictions and assign uncertainties to them.
\end{enumerate}
Table~\ref{tab:parameters} summarizes the values obtained 
for the parameters and their uncertainties. 
These values are in reasonable agreement 
with previous works~\cite{Bijker:1994yr,Bijker:2000gq}, in particular $\alpha'$ 
which corresponds to the expected slope of the projection 
of the baryon Regge trajectories into the 
$\left(\Re[s_p],J\right)$ plane where 
$s_p$ is the pole position of the resonance~\cite{Fernandez-Ramirez:2015fbq}.

\begin{table}
\caption{Parameters in Eqs.~\eqref{msq}, \eqref{msqsf} and \eqref{msqorb}. As explained in Sec.~\ref{sec:fit}, $m_n$ is fixed by the nucleon mass, $m_c$ is randomized within $1400$ and $1500$ MeV using a  uniform distribution, and $z_2$ is fixed to zero. The rest of the parameters are fitted to the states in
Tables~\ref{tab:groundstates} and~\ref{tab:excitedstates}.}
\label{tab:parameters}
\begin{ruledtabular}
\begin{tabular}{ccl}
$m_u=m_d=m_n$ & $312.973$ (fixed) & MeV \\
$m_s$ & $462.1 \pm 3.6$ & MeV \\
$m_c$ & $\phantom{00}1400$ -- $1500$ & MeV \\
$a$   & $-0.0796 \pm 0.0044$ & GeV$^2$ \\
$b_3$ & $\phantom{-}0.03708 \pm 0.00094$ & GeV$^2$ \\
$b_4$ & $-0.0122 \pm 0.0050$ & GeV$^2$ \\
$c_S$   & $\phantom{-}0.116 \pm 0.010$ & GeV$^2$ \\
$c_I$   & $\phantom{-}0.0368 \pm 0.0034$ & GeV$^2$ \\
$y_1$ & $-0.1082 \pm 0.0096$ & GeV$^2$ \\
$y_2$ & $-0.0115 \pm 0.0023$ & GeV$^2$ \\
$z_1$ & $-1.42 \pm 0.11$ & GeV$^2$ \\
$z_2$ & $\phantom{-}0$ (fixed) & GeV$^2$ \\
$\alpha'$ & $\phantom{-}1.067 \pm 0.017$ & GeV$^2$ 
\end{tabular}
\end{ruledtabular}
\end{table}
%

\subsection{Mass spectrum}
\label{sec:mass_spectrum}
In this section we present the result of a fit of the mass formula of Eqs.~\eqref{msq}, \eqref{msqsf} and \eqref{msqorb} to the charmless and single-charm baryons. 
\subsubsection{Charmless and single charm baryons}
\label{sec:charmstates}
\begin{figure}
\centering
\subfigure[\ Charmless baryons.]{
\rotatebox{0}{\scalebox{0.32}[0.32]{\includegraphics{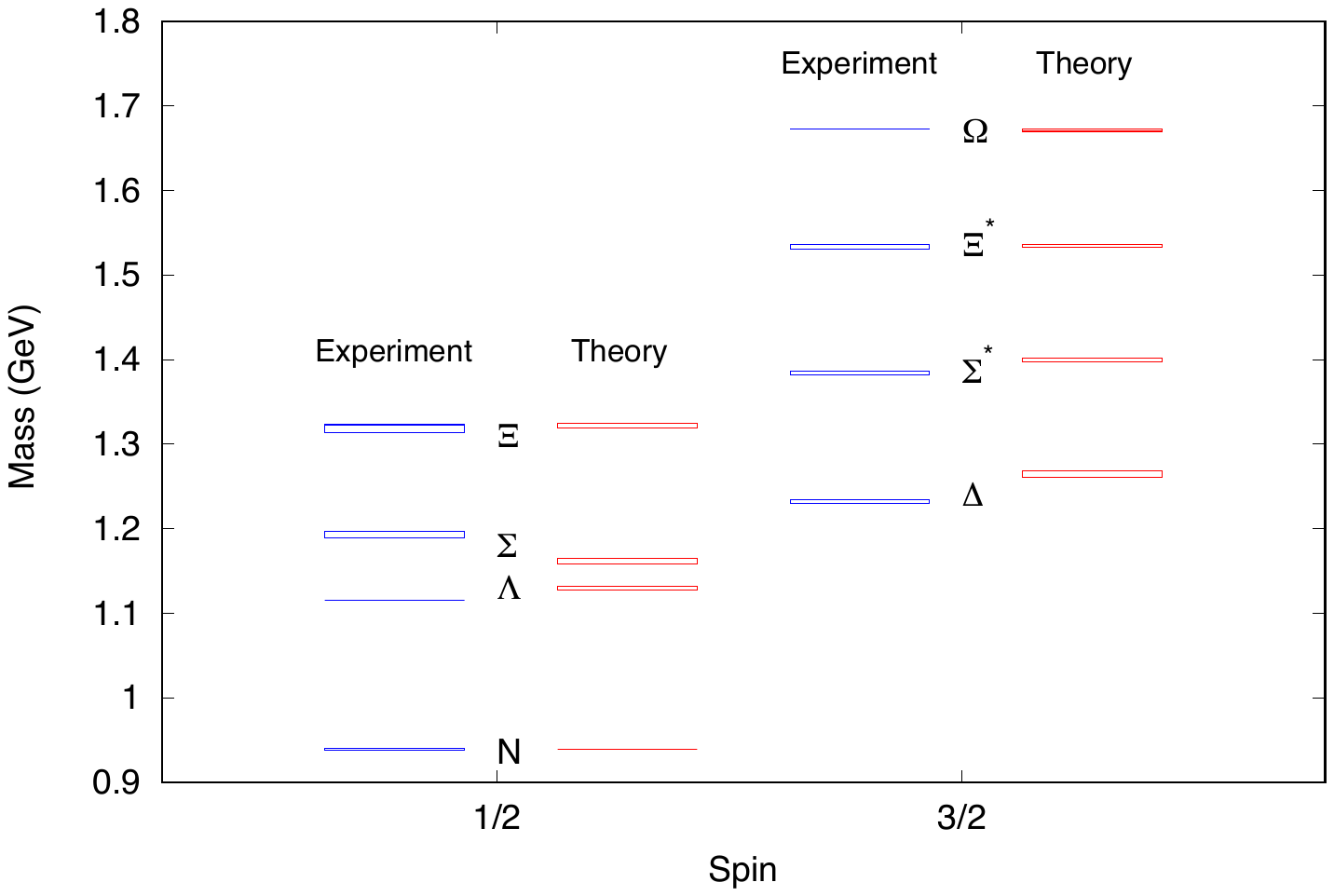}}}\label{fig:ground1}}
\subfigure[\ Single charm baryons.]{
\rotatebox{0}{\scalebox{0.32}[0.32]{\includegraphics{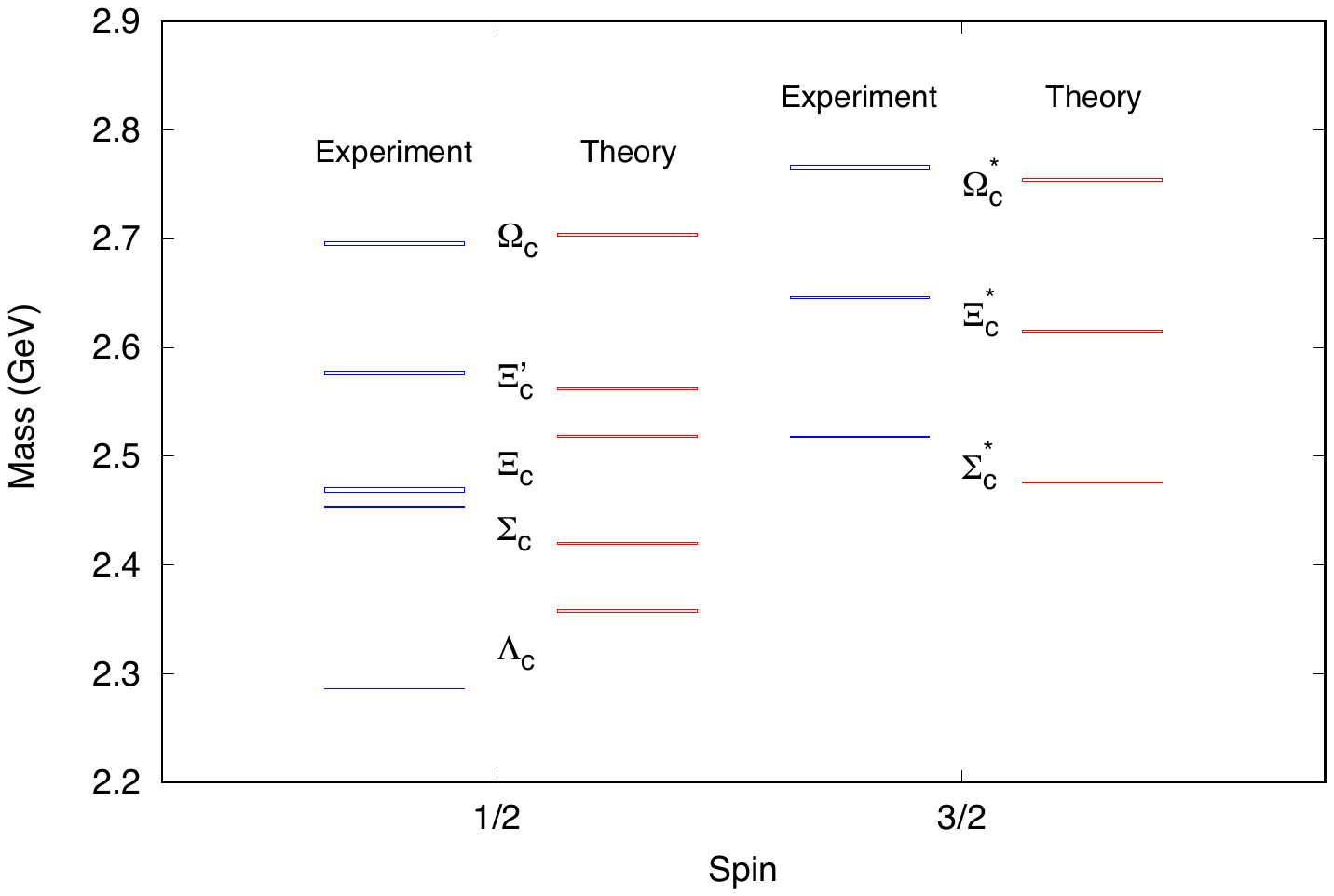}}}\label{fig:ground2}}
\caption{Fit {\it vs.} experimental data for the 
baryon ground states from Table~\ref{tab:groundstates}. 
Our fit (red boxes) {\it vs.} PDG (blue boxes) for
(a) Charmless baryon and
(b) Single charm baryon ground states.
}\label{fig:groundstates}
\end{figure}
The results for the states we fit are reported in 
Tables~\ref{tab:groundstates} and~\ref{tab:excitedstates}.
In general, we obtain a good description of the spectrum. 
The deviation of the mean value $\bar{M}^{th}$ from 
the experimental value $\bar{M}^{exp}$ are below $3\%$ for 
all the states but seven 
(four if we do not consider the fit suppressed $\Lambda$ singlets).
Without considering the uncertainties in the states, 
we quantify the quality of the description of the baryon spectrum
through
the root mean square ($\delta_{rms}$)
and the average distance ($\delta_{abs}$) between experiment and theory
\begin{equation}
\delta_{rms} =\sqrt{ \frac{1}{N_r-n_p}
\sum_{i=1}^{N_r} \left( 
\bar{M}^{exp}_{i} - \bar{M}^{th}_i \right)^2}=44~\text{MeV},
\nonumber
\end{equation}
and
\begin{equation}
\delta_{abs} = \frac{1}{N_r-n_p}\sum_{i=1}^{N_r} 
\left| \bar{M}^{exp}_{i} - \bar{M}^{th}_i  \right|=31~\text{MeV}.
\nonumber
\end{equation}

Figures~\ref{fig:ground1} and~\ref{fig:ground2} 
compare the theoretical masses to the experimental ones 
for the octet and decuplet ground
states (Table~\ref{tab:groundstates}). The width of the boxes represent the
uncertainties.
As explained in previous section, the uncertainties in 
the theoretical masses are computed
propagating exactly the uncertainties in the parameters 
as well as their correlations.
We note that charmless ground states are in general
very well reproduced --except for the $\Lambda-\Sigma$ mass splitting and the $\Delta(1232)$ mass--,
while for the single charm states
we reproduce the structure of the spectrum
but we are not that accurate when it comes to a level by level comparison.
This is expected and to accrue a high level of precision we would need a 
detailed dynamical model for the interaction among quarks.
In this work we are more concerned with the general structure of the spectrum and
this level of accuracy is enough for that purpose.
Nevertheless, in percentage all ground states but 
$\Delta(1232)$, $\Sigma(1193)$ and $\Lambda_c$
are reproduced within less than a 2\% accuracy.

The fitted excited states ($L^\pi \neq 0^+$)
are provided in Table~\ref{tab:excitedstates}. 
In general we obtain a reasonable agreement that,
except for the $\Lambda$ singlets, the $N(1680)$, $N(2250)$ and the $\Sigma(1750)$,
are within the 3\% accuracy. Hence, the charmless and single charm experimental spectrum
is reasonably reproduced.

The highest deviation ($\sim14\%$) is observed for the $\Lambda$(1405). 
The $\Lambda$(1405) is actually made out of two poles~\cite{Roca:2013cca,Mai:2014xna,Tanabashi:2018oca}
whose nature is under discussion~\cite{Fernandez-Ramirez:2015fbq,Fernandez-Ramirez:2016knc}. Some works interpret it to be a molecule~\cite{Hyodo:2007jq,Roca:2013av,Hall:2014uca} and others favor a compact three-quark state~\cite{Faustov:2015eba,Santopinto:2014opa,Engel:2012qp,Engel:2013ig,Goity:2002pu,Schat:2001xr}.
Quark-diquark models for hyperons are able to reproduce this state~\cite{Faustov:2015eba,Santopinto:2014opa}
but require dynamics that are not included in our mass formula. Therefore, it is not surprising that we are not able to properly describe this state.

The idea was to present an global overall fit including both ground-state and radially excited baryons ($N$, $\Delta$, $\Sigma$, $\Lambda$, $\Xi$ and $\Omega$ and single-charm baryons). A recent paper by Yoshida {\it et al.} on the spectrum of heavy baryons \cite{Yoshida:2015tia} shows a better agreement for low-lying charm baryons, but in this study the $N$ and $\Delta$ states were not taken into account. 

In conclusion, a good overall fit is found with a r.m.s. deviation of only 44 MeV and an average distance between experiment and theory of 31 MeV. We note that these deviations are  based on the central values and do not include the experimental error bars nor the theoretical uncertainties. Moreover, these deviations include the results for the singlet $\Lambda$ hyperons which are known not to fit in any global quark model description of baryons. With the exception of the singlet $\Lambda$ hyperons, the relative deviation is of the order of a few percent only.

\subsubsection{Double and triple charm baryons}
\label{sec:23cstates}
\begin{figure}
\rotatebox{0}{\scalebox{0.3}[0.3]{\includegraphics{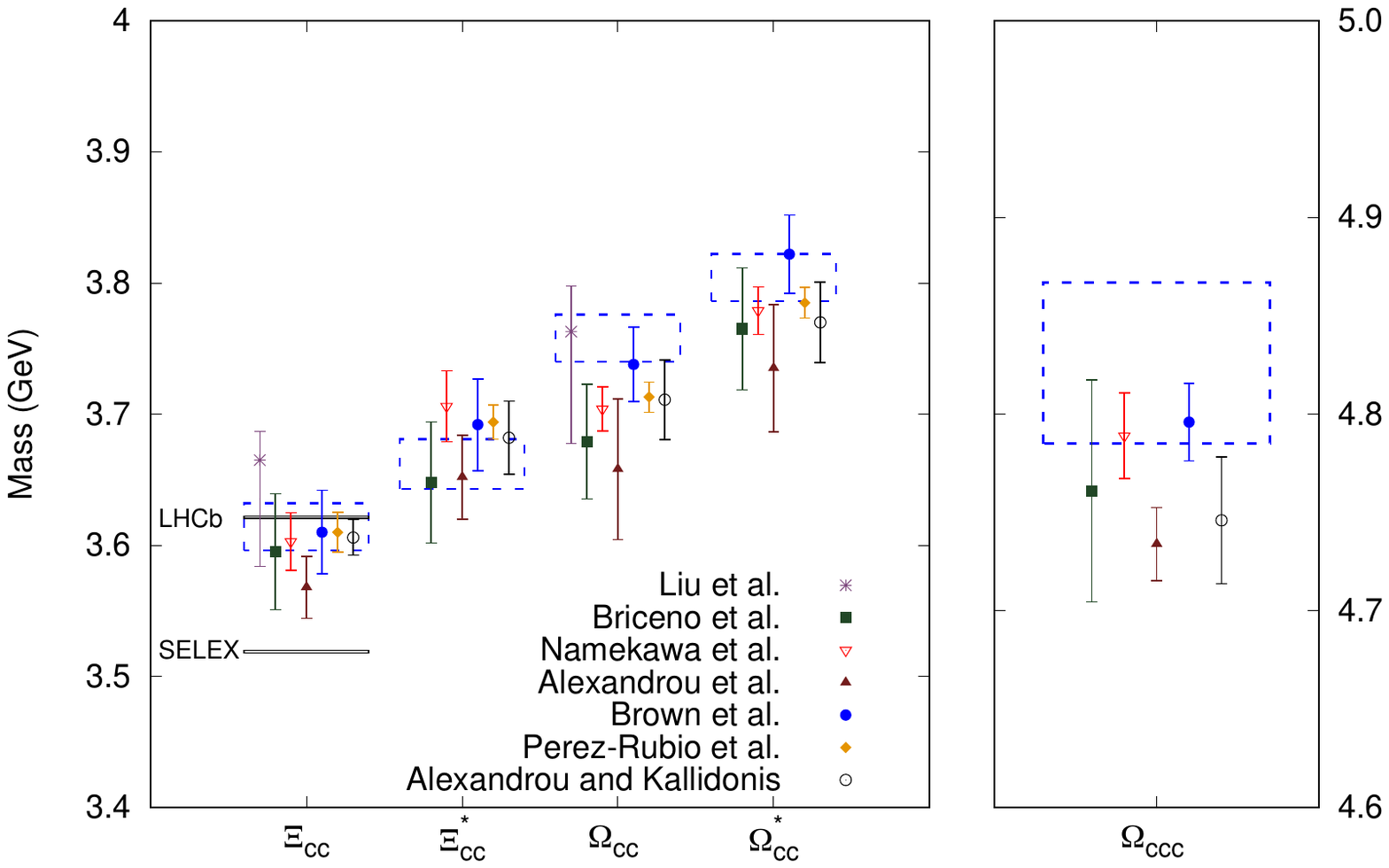}}}
\caption{
Predicted double and triple baryon ground states 
(dashed line blue boxes) 
compared to $\Xi_{cc}$ measurements by the 
SELEX~\cite{Ocherashvili:2004hi} 
and LHCb~\cite{Aaij:2017ueg} experiments
and to LQCD calculations from
Liu {\it et al.}~\cite{Liu:2009jc},
Brice\~no {\it et al.}~\cite{Briceno:2012wt},
Namekawa {\it et al.}~\cite{Namekawa:2013vu},
Alexandrou {\it et al.}~\cite{Alexandrou:2014sha},
Brown {\it et al.}~\cite{Brown:2014ena},
P\'erez-Rubio {\it et al.}~\cite{Bali:2015lka},
and Alexandrou and Kallidonis~\cite{Alexandrou:2017xwd}.
}\label{fig:ground3}
\end{figure}

\begin{table*}
\caption{As Table~\ref{tab:groundstates}, but for predicted 
ground state double and triple charm baryons. 
In Fig.~\ref{fig:ground3} we compare our results 
to those from experiments and LQCD calculations.} \label{tab:predictedgroundstates}
\begin{ruledtabular}
\begin{tabular}{lccccrrcccccc}
Name & $[f]$ & $(\alpha, \beta, \gamma)$ & $(\lambda, \mu)$ 
& $I$ & $Y$ & $Z$ & $S^{\cal P}=J^P$ & $(N_n,N_s,N_c)$ & $M^{th}$ (MeV) \\
\hline
$\Xi_{cc}$ & $[3]$ &  $(1,1,0)$ & $(1,0)$ 
& $\frac{1}{2}$ & $\frac{1}{3}$ & $-\frac{5}{4}$ & 
$\frac{1}{2}^+$ & $(1,\: 0,\: 2)$ 
& $3614 \pm  18$  \\
$\Omega_{cc}$ & $[3]$ & $(1,1,0)$ & $(1,0)$ 
& $0$ & $-\frac{2}{3}$ & $-\frac{5}{4}$ & $\frac{1}{2}^+$ & $(0,\: 1,\: 2)$ 
& $3758 \pm 18$  \\   
$\Xi^*_{cc}$ & $[3]$ & $(3,0,0)$ & $(1,0)$ 
& $\frac{1}{2}$ & $\frac{1}{3}$ & $-\frac{5}{4}$ & 
$\frac{3}{2}^+$ & $(1,\: 0,\: 2)$ 
& $3662 \pm 19$ \\
$\Omega^*_{cc}$ & $[3]$ & $(3,0,0)$ & $(1,0)$ 
& $0$ & $-\frac{2}{3}$ & $-\frac{5}{4}$ & $\frac{3}{2}^+$ & $(0,\: 1,\: 2)$ 
& $3804 \pm 18$  \\
$\Omega_{ccc}$ & $[3]$ & $(3,0,0)$ & $(0,0)$ 
& $0$ & $0$ & $-\frac{9}{4}$ & $\frac{3}{2}^+$ & $(0,\: 0,\: 3)$ 
& $4826 \pm 41$ 
\end{tabular}
\end{ruledtabular}
\end{table*}
In our fits we do not include existing experimental information 
on double charm states.
There is one measurement of $\Xi_{cc}$ reported 
by the SELEX collaboration
at $3518.9 \pm 0.9$ MeV~\cite{Mattson:2002vu,Ocherashvili:2004hi}. 
However, we do not consider it reliable enough 
to be included in our fits 
because it is listed by the PDG~\cite{Tanabashi:2018oca} 
as a one-star state, 
and has not been observed by either 
the BaBar~\cite{Aubert:2006qw},
Belle~\cite{Chistov:2006zj,Kato:2013ynr} 
or LHCb~\cite{Aaij:2013voa} collaborations.
Recently, the LHCb collaboration reported the measurement
of a double-charm $\Xi_{cc}$ state 
at $3621.40 \pm 0.78$ MeV~\cite{Aaij:2017ueg},
at odds with the SELEX measurement. 
In the fit we do not include this state either 
as further experimental confirmation is needed. 
In Table~\ref{tab:predictedgroundstates} 
we provide our predictions for 
double and triple charm ground states, 
($\Xi_{cc}$, $\Xi_{cc}^*$, $\Omega_{cc}$, 
$\Omega_{cc}^*$ and $\Omega_{ccc}$)
and in Fig.~\ref{fig:ground3} we compare them to the 
experimental data and LQCD
computations~\cite{Liu:2009jc,Briceno:2012wt,Namekawa:2013vu,Alexandrou:2014sha,Brown:2014ena,Bali:2015lka,Alexandrou:2017xwd}.
Our results are in good agreement with the LHCb measurement and the
LQCD computations. 
Consequently, our choice to fix $z_2=0$ 
in the mass formula provides sensible 
results for the double and triple charm states. 
Therefore, we expect that
our prediction of the hidden charm pentaquark spectrum, 
in particular the structure and the ground states, 
should be reasonable within uncertainties.
\section{Pentaquarks}
\label{sec:pentaquarks}
As for all multiquark systems, the pentaquark wave function contains  contributions corresponding to the spatial degrees of freedom  
and the internal degrees of freedom of color, flavor and spin. 
In order to classify the corresponding states, we shall make use as 
much as possible of symmetry principles without, for the moment, 
introducing any explicit dynamical model. In the construction of the 
classification scheme we are guided by two conditions: (i) the pentaquark wave function should be antisymmetric under any permutation of the four quarks, and (ii) as all physical states, it should be a color singlet. In addition, we restrict ourselves to flavor multiplets containing $uudc\bar{c}$ configurations without orbital excitations, {\it i.e.} with $L^{\pi}=0^+$, and with spin and parity $J^P=S^{\cal P}=3/2^-$.

In order to discuss the symmetry properties of the pentaquark wave function we start by introducting the notation for the symmetry properties of the four-quark subsystem. The permutation symmetry for four-quark states is characterized by the $S_4$ Young tableaux $[4]$, $[31]$, $[22]$, $[211]$ and $[1111]$ or, equivalently, by the irreducible representations of the tetrahedral group ${\cal T}_d$ (which is isomorphic to $S_4$) as $A_1$, $F_2$, $E$, $F_1$ and $A_2$, respectively 
\ba
\begin{array}{rcccl}
\, [4]    &\sim& A_1 && \mbox{singlet}, \\
\, [31]   &\sim& F_2 && \mbox{triplet}, \\
\, [22]   &\sim& E   && \mbox{doublet}, \\
\, [211]  &\sim& F_1 && \mbox{triplet}, \\
\, [1111] &\sim& A_2 && \mbox{singlet}.
\end{array} 
\label{su4td} 
\ea
The total pentaquark wave function has to be a $[222]$ color-singlet state. Since the color wave function of the antiquark is a $[11]$ anti-triplet, the color wave function of the four-quark configuration has to be a $[211]$ triplet with $F_1$ symmetry under ${\cal T}_d$. 
Since the total $q^4$ wave function has to be antisymmetric ($A_2$), the 
orbital-spin-flavor part is a $[31]$ state with $F_2$ symmetry 
\ba
\psi &=& 
\left[ \psi^{\rm c}_{F_1} \times \psi^{\rm osf}_{F_2} \right]_{A_2} ~,
\label{pentaquarkwf}
\ea
where the subindices refer to the symmetry properties of the four-quark subsystem under permutation. Moreover, it is assumed that the orbital part of the pentaquark wave function corresponds to the ground state (without orbital excitations) with $A_1$ symmetry. As a consequence, the spin-flavor part is a $[f]=[31]$ state with $F_2$ symmetry. The pentaquark wave function of Eq.~\eqref{pentaquarkwf} can be expanded explicitly into its color and spin-flavor parts by using the tensor couplings under the permutation group $S_4 \sim {\cal T}_d$ (Clebsch-Gordan coefficients) \cite{Stancu:1991rc} 
\begin{equation}
\begin{split}
\psi =& \, \psi^{\rm o}_{A_1} \psi^{\rm csf}_{A_2} \\
  =& \frac{1}{\sqrt{3}} \left[ \psi^{\rm o}_{A_1} \left( 
  \psi^{\rm c}_{F_{1\lambda}} \psi^{\rm sf}_{F_{2\rho}} 
- \psi^{\rm c}_{F_{1\rho}}    \psi^{\rm sf}_{F_{2\lambda}} 
+ \psi^{\rm c}_{F_{1\eta}}    \psi^{\rm sf}_{F_{2\eta}} \right) \right] ~. 
\end{split}
\label{wfneg}
\end{equation}
\subsection{Four-quark states}
\label{sec:4qstates}
\begin{figure}
\centering
\setlength{\unitlength}{0.4pt}
\begin{picture}(500,400)(-40,0)
\thicklines
\put(230,365){$uudc\bar{c}$}
\put( 50,350) {\line(1,0){300}}
\put(100,300) {\line(1,0){200}}
\put(150,250) {\line(1,0){100}}
\put(200,200) {\line(1,1){150}}
\put(150,250) {\line(1,1){100}}
\put(100,300) {\line(1,1){ 50}}
\put(200,200) {\line(-1,1){150}}
\put(250,250) {\line(-1,1){100}}
\put(300,300) {\line(-1,1){ 50}}
\put(250,350){\circle*{10}}
\multiput( 50,350)(100,0){4}{\circle*{5}}
\multiput(100,300)(100,0){3}{\circle*{5}}
\multiput(150,250)(100,0){2}{\circle*{5}}
\put(200,200){\circle*{5}}
\put(230,165){$uudc\bar{c}$}
\put(150,150) {\line(1,0){100}}
\put(100,100) {\line(1,0){200}}
\put(150, 50) {\line(1,0){100}}
\put(100,100) {\line(1,1){ 50}}
\put(150, 50) {\line(1,1){100}}
\put(250, 50) {\line(1,1){ 50}}
\put(150, 50) {\line(-1,1){ 50}}
\put(250, 50) {\line(-1,1){100}}
\put(300,100) {\line(-1,1){ 50}}
\put(250,150){\circle*{10}}
\multiput(150,150)(100,0){2}{\circle*{5}}
\multiput(100,100)(100,0){3}{\circle*{5}}
\multiput(150, 50)(100,0){2}{\circle*{5}}
\put(200,100){\circle{10}}
\put(375,350) {$nnnc\bar{c}$}
\put(375,300) {$nnsc\bar{c}$}
\put(375,250) {$nssc\bar{c}$}
\put(375,200) {$sssc\bar{c}$}
\put(375,150) {$nnnc\bar{c}$}
\put(375,100) {$nnsc\bar{c}$}
\put(375, 50) {$nssc\bar{c}$}
\put(  0,350) {$P_c^{\Delta}$}
\put(  0,300) {$P_c^{\Sigma^*}$}
\put(  0,250) {$P_c^{\Xi^*}$}
\put(  0,200) {$P_c^{\Omega}$}
\put(-35,150) {$P_c^{N} \equiv P_c$}
\put(-25,100) {$P_c^{\Sigma}/P_c^{\Lambda}$}
\put(  0, 50) {$P_c^{\Xi}$}
\end{picture}
\caption{Pentaquark decuplet and octet notation for hidden charm pentaquarks according to 
Eqs.~\eqref{qqqqflavor10}, \eqref{qqqqflavor8} and \eqref{penta3}. The notation $n$ refers to $u$ and/or $d$ quarks.}
\label{notation}
\end{figure}
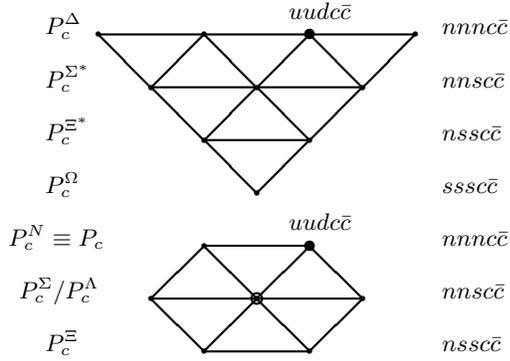

First we construct all $q^4$ spin-flavor multiplets containing a $uudc$ state to which we couple the antiquark $\bar{q}=\bar{c}$ to obtain all pure $uudc\bar{c}$ states. The spin-flavor states of four-quark systems can be obtained by taking the products of the representations of the quarks 
\begin{equation}
\begin{split}
[1]_8 \;\otimes\; [1]_8 \;\otimes\; [1]_8 \;\otimes\; [1]_8 \;=\; 
[4]_{330} \;\oplus\; 3 \, [31]_{630} \\
\;\oplus\; 2 \, [22]_{336} \;\oplus\; 3 \, [211]_{378} 
\;\oplus\; [1111]_{70} ~. 
\end{split}\label{qqqqsu8}
\end{equation}
A complete classification of four-quark states involves the analysis of the flavor and spin content. In Table~\ref{sfqqqq} we show the result for the spin-flavor configuration $[f]=[31]$ with $F_2$ symmetry. There are four different flavor multiplets $[g]=[4]$, $[31]$, $[22]$ and $[211]$. For pentaquark states with $J^P=3/2^-$ without orbital excitations the allowed values of the spin of the four-quark system are $S=(g'_1-g'_2)/2=1$ and $2$. In the last column, we show the allowed spin-flavor configurations where $\phi$ and $\chi$ denote the flavor and spin wave functions, respectively. The explicit form of the spin wave functions of pentaquark states with $J^P=3/2^-$ is presented in Appendix~\ref{app:spin}. 
\begin{table}
\caption[]{Spin-flavor decomposition of $q^4$ states. 
The subindices denote the dimensions of the multiplets. }
\label{sfqqqq}
\begin{ruledtabular}
\begin{tabular}{ccccccc}
$SU_{\rm sf}(8)$ &$\supset$& $SU_{\rm f}(4)$ 
&$\otimes$& $SU_{\rm s}(2)$ & \\
\noalign{\smallskip}
$[f]$ &$\supset$& $[g]$ &$\otimes$& $[g']$ & $\psi^{\rm sf}_{F_2}$ \\
\noalign{\smallskip}
\noalign{\smallskip}
\hline
\noalign{\smallskip}
$[31]_{630}$ & & $[4]_{35}$   & $\otimes$ & $[31]_{3}$ & $\left[ \phi_{A_1} \times \chi_{F_2} \right]_{F_2}$ \\
             & & $[31]_{45}$  & $\otimes$ & $[4]_{5}$  & $\left[ \phi_{F_2} \times \chi_{A_1} \right]_{F_2}$ \\
             & &              &           & $[31]_{3}$ & $\left[ \phi_{F_2} \times \chi_{F_2} \right]_{F_2}$ \\ 
             & &              &           & $[22]_{1}$ & \\ 
             & & $[22]_{20}$  & $\otimes$ & $[31]_{3}$ & $\left[ \phi_{E}   \times \chi_{F_2} \right]_{F_2}$ \\
             & & $[211]_{15}$ & $\otimes$ & $[31]_{3}$ & $\left[ \phi_{F_1} \times \chi_{F_2} \right]_{F_2}$ \\ 
             & &              &           & $[22]_{1}$ &
\end{tabular}
\end{ruledtabular}
\end{table}
Since the current interest is in four-quark states with one charm quark, it is convenient to make a decomposition of four into three flavors according to $SU_{\rm f}(4) \supset SU_{\rm f}(3) \otimes U_{\rm Z}(1)$ (see Table~\ref{fqqqq}). The $SU(3)$ representations $[h]=[4]$, $[31]$, $[22]$ and $[211]$ correspond to $uds$ states with $Z=\frac{3}{4}B-C=1$; $[3]$, $[21]$ and $[111]$ to configurations with one charm quark and $Z=0$; $[2]$ and $[11]$ to states with two charm quarks and $Z=-1$; $[1]$ to states with three charm quarks and $Z=-2$, and $[0]$ to a state of four charm quarks with $Z=-3$. Table~\ref{fqqqq} shows that the states with one charm quark have $Z=0$ and belong to either an $SU(3)$ decuplet ($[h]=[3]$), octet ($[21]$) or singlet ($[111]$). Since the flavor singlet corresponds to a $udsc$ configuration, the only $SU(3)$ flavor multiplets that contain a $uudc$ state are the decuplet and the octet (see Fig.~\ref{notation}). 
\begin{table*}
\caption[]{$SU_{\rm f}(4) \supset SU_{\rm f}(3) \otimes U_{\rm Z}(1)$ flavor classification of four-quark states (here $q=u$, $d$, $s$).}
\label{fqqqq}
\begin{ruledtabular}
\begin{tabular}{ccccccccccc}
$SU_{\rm f}(4)$ &$\supset$& $SU_{\rm f}(3)$ && && && && \\
\noalign{\smallskip}
$[g]$ &$\supset$& $[h]$ && && && && \\
\noalign{\smallskip}
\hline
\noalign{\smallskip}
$[4]_{35}$   &$\supset$& $[4]_{15}$ &$\oplus$& $[3]_{10}$ &$\oplus$& $[2]_{6}$ &$\oplus$& $[1]_{3}$ &$\oplus$& $[0]_{1}$ \\
\noalign{\smallskip}
$[31]_{45}$  &$\supset$& $[31]_{15}$ &$\oplus$& $[3]_{10} \oplus [21]_{8}$ &$\oplus$& $[2]_{6} \oplus [11]_{3}$ &$\oplus$& $[1]_{3}$ && \\
\noalign{\smallskip}
$[22]_{20}$  &$\supset$& $[22]_{6}$ &$\oplus$& $[21]_{8}$ &$\oplus$& $[2]_{6}$ && && \\
\noalign{\smallskip}
$[211]_{15}$ &$\supset$& $[211]_{3}$ &$\oplus$& $[21]_{8} \oplus [111]_{1}$ &$\oplus$& $[11]_{3}$ && && \\
\noalign{\smallskip}
$[1111]_{1}$ &$\supset$& && $[111]_{1}$ && && && \\
\noalign{\smallskip}
\hline
\noalign{\smallskip}
&& $qqqq$ && $qqqc$ && $qqcc$ && $qccc$ && $cccc$ \\
&& $Z=1$ && $Z=0$ && $Z=-1$ && $Z=-2$ && $Z=-3$ \\ 
\end{tabular}
\end{ruledtabular}
\end{table*}
In summary, the four-quark flavor configurations can be expressed in terms of the quantum numbers as 
\begin{equation}
\left| \phi_{t}(q^4) \right> = \left| [g],[h],I,I_3,Y,Z \right>_t ~. 
\end{equation}
There is a one-to-one correspondence between the $SU(4)$ flavor labels $[g]$ of the four-quark system and the label $t$ of the tetrahedral group (see Eq.~\eqref{su4td}). The possible $uudc$ states are given by 
\begin{equation}
\begin{split}
\left| \phi_{A_1}(uudc) \right> &= \left| [4],[3],\frac{3}{2},\frac{1}{2},1,0 \right>_{A_1} ~,\\
\left| \phi_{F_2}(uudc) \right> &= \left| [31],[3],\frac{3}{2},\frac{1}{2},1,0 \right>_{F_2} ~, 
\end{split}\label{qqqqflavor10}
\end{equation}
for the decuplet, and  
\begin{equation}
\begin{split}
\left| \phi_{F_2}(uudc) \right> &= \left| [31],[21],\frac{1}{2},\frac{1}{2},1,0 \right>_{F_2} ~,\\
\left| \phi_{E}(uudc)   \right> &= \left| [22],[21],\frac{1}{2},\frac{1}{2},1,0 \right>_{E} ~,\\
\left| \phi_{F_1}(uudc) \right> &= \left| [211],[21],\frac{1}{2},\frac{1}{2},1,0 \right>_{F_1} ~, 
\end{split}\label{qqqqflavor8}
\end{equation}
for the octet. 
%
\subsection{Pentaquark states}
\label{sec:PentaquarkStates}
The $SU(3)$ flavor multiplets which contain $uudc\bar{c}$ pentaquark states are obtained by taking the direct product of the four-quark $uudc$ states of Eqs.~(\ref{qqqqflavor10}) and~(\ref{qqqqflavor8}), and the antiquark state with charm $\bar{c}$ of Eq.~\eqref{antiquarkwf}
\begin{equation}
\left| \phi(\bar{c}) \right> = - \left| [111],[111],0,0,0,\frac{3}{4} \right> ~. \label{antic}
\end{equation}
Inspection of the $SU(3)$ flavor couplings 
\begin{equation}
\begin{split}
\, [3]_{q^4} \otimes [111]_{\bar{q}} &= [411] ~, \\
\, [21]_{q^4} \otimes [111]_{\bar{q}} &= [321] ~,
\end{split}
\end{equation}
shows that the $SU(3)$ flavor multiplet which contains a $uudc\bar{c}$ pentaquark can either be a decuplet with $[h]=[411] \equiv [3]$ or an octet $[321] \equiv [21]$. 

The total flavor wave function of the pentaquark configurations obtained by coupling the four-quark states and the antiquark state at the level of $SU(3)$ flavor is given by 
\begin{equation}
\left| \underbrace{[f],[g],[h]}_{q^4},
\underbrace{[f],[g],[h]}_{\bar{q}};
\underbrace{[h],I,I_3,Y,Z}_{q^4\bar{q}} \right> ~. 
\label{wf3}
\end{equation}
For the case of $uudc\bar{c}$ states this wave function factorizes into a product of a four-quark state given by Eqs.~\eqref{qqqqflavor10} 
and~\eqref{qqqqflavor8}, and the antiquark state of Eq.~\eqref{antic}  
\begin{equation}
\left| \phi_t(uudc\bar{c}) \right> = \left| \phi_t(uudc) \right> 
\left| \phi(\bar{c}) \right> ~. 
\label{penta3}
\end{equation}
In Table~\ref{flavorwf} of Appendix~\ref{app:flavor1} we present the explicit form of these flavor wave functions in terms of the individual quarks. Note that they correspond to pure $uudc\bar{c}$ states, {\it i.e.} not admixed with $uudu\bar{u}$, $uudd\bar{d}$ or $uuds\bar{s}$ configurations. 

However, in order to apply the mass formula of Eq.~\eqref{msq} to the pentaquark mass spectrum, we have to express the pentaquark wave functions in terms of the $SU(4)$ flavor couplings. If one couples, just as for the three-quark baryons discussed in Sec.~\ref{sec:3qbaryons}, the flavor states at the level of $SU(4)$ flavor the pentaquark wave function is given by 
\begin{equation}
\left| \underbrace{[f],[g]}_{q^4},
\underbrace{[f],[g]}_{\bar{q}};
\underbrace{[g],[h],I,I_3,Y,Z}_{q^4\bar{q}} \right> ~.
\end{equation}
In the following, the labels that are the same for all states of interest, $[f]_{q^4}=[31]$, $[f]_{\bar{q}}=[1^7]$, and $[g]_{\bar{q}}=[111]$ will not be written explicitly anymore. Moreover, since there is a one-to-one correspondence between the $SU(4)$ flavor labels $[g]$ of the four-quark system and the label $t$ of the tetrahedral group 
(see Eq.~\eqref{su4td}), we will use the latter as a subindex to indicate the flavor symmetry of the four-quark subsystem. In conclusion, in the four-flavor coupling scheme the flavor wave functions of the pentaquark states are labeled as
\begin{equation}
\left| \phi_{t}(q^4\bar{q}) \right> = \left| [g],[h],I,I_3,Y,Z \right>_t ~.  
\end{equation}
The allowed four-flavor wave functions of the  $q^4 \bar{q}$ pentaquark states can be obtained by taking the direct product of the Young tableaux of the four-quark states, $[g]=[4]$, $[31]$, $[22]$ and $[211]$ (see Table~\ref{sfqqqq}), and that of the antiquark $[111]$ 
\ba
\begin{array}{lcrcl}
A_1 &:& [4]_{q^4} \otimes [111]_{\bar{q}} &=& [511] \oplus [4111] ~, \\
F_2 &:& [31]_{q^4} \otimes [111]_{\bar{q}} &=& [421] \oplus [4111] \oplus [3211] ~, \\
E &:& [22]_{q^4} \otimes [111]_{\bar{q}} &=& [331] \oplus [3211] ~, \\
F_1 &:& [211]_{q^4} \otimes [111]_{\bar{q}} &=& [322] \oplus [3211] \oplus [2221] ~. 
\end{array} 
\nonumber\\
\mbox{}
\ea
As a result, we find that the configurations with the same quantum numbers $\left| [h],I,I_3,Y,Z \right>$ as the $uudc\bar{c}$ states of Eq.~\eqref{wf3} are given by 
\ba
\, [g]_t &=& [511]_{A_1} , \; [4111]_{A_1} ~, 
\nonumber\\
&& [421]_{F_2} , \; [4111]_{F_2} ~, 
\label{decuplet4}
\ea
for the decuplet with $[h]=[411] \equiv [3]$ and $\left| I,I_3,Y,Z \right>=\left| \frac{3}{2},\frac{1}{2},1,\frac{3}{4} \right>$, and
\ba
\, [g]_t &=& [421]_{F_2} ~, \; [3211]_{F_2} ~, 
\nonumber\\
&& [331]_{E} ~, \; [3211]_{E} ~, 
\nonumber\\
&& [322]_{F_1} ~, \; [3211]_{F_1} ~, 
\label{octet4}
\ea
for the octet with $[h]=[321] \equiv [21]$ and $\left| I,I_3,Y,Z \right>=\left| \frac{1}{2},\frac{1}{2},1,\frac{3}{4} \right>$. 
We note, that the classification of hidden-charm pentaquark states in Eqs.~(\ref{decuplet4}) and (\ref{octet4}) is related to the $SU(4)$ classification of Ref.~\cite{Wu:2004wg} via the correspondence of the $\overline{60}$-plet with $[g]_t=[331]_{E}$, and the 140-plet with  $[g]_t=[421]_{F_2}$.

It is important to note that in this coupling scheme the states of Eqs.~\eqref{decuplet4} and~\eqref{octet4} do not correspond to pure $uudc\bar{c}$ states but contain admixtures with $uudu\bar{u}$, $uudd\bar{d}$ or $uuds\bar{s}$ configurations. 
For example, the decuplet configurations with $A_1$ symmetry of Eq.~\eqref{decuplet4} can be expanded into a basis of product states of four-quark states and an antiquark state by inserting the appropriate $SU(4)$ Clebsch-Gordan coefficients~\cite{Alex:2010wi} to give
\ba
&& \left| [511],[411],\frac{3}{2},\frac{1}{2},1,
\frac{3}{4} \right>_{A_1} 
\nonumber\\
&& \hspace{1cm} = 
-\sqrt{\frac{6}{7}} \left| [4],[3],\frac{3}{2},
\frac{1}{2},1,0 \right>_{A_1} \left| \phi(\bar{c}) \right>  
\nonumber\\
&& \hspace{1.2cm} +\frac{1}{\sqrt{42}} \left| [4],[4],\frac{3}{2},
\frac{1}{2},\frac{1}{3},1 \right>_{A_1} 
\left| \phi(\bar{s}) \right>  
\nonumber\\
&& \hspace{1.2cm} +\frac{1}{\sqrt{21}} \left| [4],[4],2,0,
\frac{4}{3},1 \right>_{A_1} 
\left| \phi(\bar{d}) \right>  
\nonumber\\
&& \hspace{1.2cm} +\frac{1}{\sqrt{14}} \left| [4],[4],2,1,
\frac{4}{3},1 \right>_{A_1} \left| \phi(\bar{u}) \right> ~, \label{A1d1}
\ea
and
\ba
&& \left| [4111],[411],\frac{3}{2},\frac{1}{2},1,
\frac{3}{4} \right>_{A_1} 
\nonumber\\ 
&& \hspace{1cm} = 
\frac{1}{\sqrt{7}} \left| [4],[3],\frac{3}{2},
\frac{1}{2},1,0 \right>_{A_1} \left| \phi(\bar{c}) \right>  
\nonumber\\
&& \hspace{1.2cm} +\frac{1}{\sqrt{7}} \left| [4],[4],\frac{3}{2},\frac{1}{2},\frac{1}{3},1 \right>_{A_1} \left| \phi(\bar{s}) \right>  
\nonumber\\
&& \hspace{1.2cm} +\sqrt{\frac{2}{7}} \left| [4],[4],2,0,\frac{4}{3},1 \right>_{A_1} \left| \phi(\bar{d}) \right>  
\nonumber\\
&& \hspace{1.2cm} +\sqrt{\frac{3}{7}} \left| [4],[4],2,1,\frac{4}{3},1 \right>_{A_1} \left| \phi(\bar{u}) \right> ~, \label{A1d2}
\ea
corresponding to linear combinations of $uudc\bar{c}$, $uuds\bar{s}$, $uudd\bar{d}$ and $uudu\bar{u}$ states. Eqs.~(\ref{A1d1}) and (\ref{A1d2}) can be combined to express the $uudc\bar{c}$ configuration with $A_1$ symmetry in terms of the $SU(4)$ flavor basis as 
\ba 
\left| \phi_{A_1}(uudc\bar{c}) \right> &=& 
\left| [4],[3],\frac{3}{2},\frac{1}{2},1,0 \right>_{A_1} 
\left| \phi(\bar{c}) \right>  
\nonumber\\
&=& -\sqrt{\frac{6}{7}} \left| [511],[411],\frac{3}{2},\frac{1}{2},1,\frac{3}{4} \right>_{A_1} 
\nonumber\\
&& +\frac{1}{\sqrt{7}} \left| [4111],[411],\frac{3}{2},\frac{1}{2},1,\frac{3}{4} \right>_{A_1} ~.
\label{wfA1d}
\ea
In a similar fashion we can express the $uudc\bar{c}$ configuration with $F_2$, $E$ and $F_1$ symmetry in the four-flavor $SU(4)$ basis (for more details see Appendix~\ref{app:flavor2}). As a result, we find 
\ba
\left| \phi_{F_2}(uudc\bar{c}) \right> &=& 
-\sqrt{\frac{2}{3}} \left| [421],[411],\frac{3}{2},\frac{1}{2},1,\frac{3}{4} \right>_{F_2}
\nonumber\\
&&-\frac{1}{\sqrt{3}} \left| [4111],[411],\frac{3}{2},\frac{1}{2},1,\frac{3}{4} \right>_{F_2} 
\label{wfF2d}
\ea
for the decuplet with $F_2$ symmetry, 
\ba
\left| \phi_{F_2}(uudc\bar{c}) \right> &=& 
-\sqrt{\frac{5}{6}} \left| [421],[321],\frac{1}{2},\frac{1}{2},1,\frac{3}{4} \right>_{F_2}
\nonumber\\
&&+\frac{1}{\sqrt{6}} \left| [3211],[321],\frac{1}{2},\frac{1}{2},1,\frac{3}{4} \right>_{F_2} 
\label{wfF2o}
\ea
for the octet with $F_2$ symmetry,
\ba
\left| \phi_{E}(uudc\bar{c}) \right> &=& 
-\frac{\sqrt{3}}{2} \left| [331],[321],\frac{1}{2},\frac{1}{2},1,\frac{3}{4} \right>_{E}
\nonumber\\
&& -\frac{1}{2} \left| [3211],[321],\frac{1}{2},\frac{1}{2},1,\frac{3}{4} \right>_{E} 
\label{wfEo}
\ea
for the octet with $E$ symmetry, and
\ba
\left| \phi_{F_1}(uudc\bar{c}) \right> &=& 
-\frac{1}{\sqrt{2}} \left| [322],[321],\frac{1}{2},\frac{1}{2},1,\frac{3}{4} \right>_{F_1} 
\nonumber\\
&& +\frac{1}{\sqrt{2}} \left| [3211],[321],\frac{1}{2},\frac{1}{2},1,\frac{3}{4} \right>_{F_1}
\label{wfF1o}
\ea
for the octet with $F_1$ symmetry.

\subsection{Mass spectrum}
\label{5qspectrum}
In this section we calculate the spectrum of the
ground-state
$qqqc\bar{c}$ pentaquarks with $q=u, d, s$,
{\it i.e.} the pentaquarks 
depicted in Fig.~\ref{notation}
without orbital excitations ($L^{\pi}=0^+$) and 
that have spin and parity 
$J^P=S^{\cal P}=3/2^-$.
As we showed in Sec.~\ref{sec:PentaquarkStates} for the case of 
$uudc\bar{c}$ pentaquarks,
Eqs.~\eqref{wfA1d}-\eqref{wfF1o}, 
the states with a well-defined flavor content are 
a superposition of different $SU_{\rm f}(4)$ representations.
Consequently, the $b_4$ term in the 
mass formula of Eq.~\eqref{msqsf} is modified as follows
\begin{align}
b_4 \left[ C_{2SU_{\rm f}(4)}-\frac{39}{8} \right]
\rightarrow &\,
b_4 \left\{ x_1^2 
\left[ C_{2SU_{\rm f,1}(4)}-\frac{39}{8} \right] 
\right. \nonumber \\
&\, \, \, \, + \left. x_2^2 
\left[ C_{2SU_{\rm f,2}(4)}-\frac{39}{8} \right] \right\},
\end{align}
where $x_1$ and $x_2$ are the expansion 
coefficients that appear in Eqs.~\eqref{wfA1d}-\eqref{wfF1o},
\textit{e.g.} $x_1=-/\sqrt{2}$ 
and $x_2=1/\sqrt{2}$ in Eq.~\eqref{wfF1o} for the
$\left| \phi_{F_1}(uudc\bar{c}) \right>$ state.
As discussed in Sec.~\ref{sec:pentaquarks}, 
the spin-flavor part of all ground-state pentaquark states is 
a $[f]=[31]$ state with $F_2$ symmetry. 
\begin{table}
\caption{Hidden charm $qqqc\bar{c}$ pentaquark states with $L^{\pi}=0^+$, 
$J^P=S^{\cal P}=3/2^-$ and $Z=3/4$ for flavor decuplet (top) 
and octet (bottom). The $P_c^N$ candidates reported by LHCb 
have $4380$ and $4450$ MeV
with opposite parities and uncertain values for $J$
\cite{Aaij:2015tga,Jurik:2016bdm}. 
\label{tab:pentaquarkstates}}
\begin{ruledtabular}
\begin{tabular}{ccrrcl}
State & Name & $I$ & $Y$ & $(N_n,N_s,N_c)$ & $M^{th}$ (MeV) \\
\hline
$\left[ \phi_{A_1} \times \chi_{F_2} \right]_{F_2}$ 
& $P_c^{\Delta}$   & $\frac{3}{2}$ & $ 1$ & $(3,0,2)$ & 
$3855\pm  54$ \\
& $P_c^{\Sigma^*}$ & $1$           & $ 0$ & $(2,1,2)$ & 
$4024 \pm 52 $ \\
& $P_c^{\Xi^*}$    & $\frac{1}{2}$ & $-1$ & $(1,2,2)$ & 
$4193 \pm  49  $ \\
& $P_c^{\Omega}$   & $0$           & $-2$ & $(0,3,2)$ & 
$4364 \pm 46 $ \\
$\left[ \phi_{F_2} \times \chi_{A_1} \right]_{F_2}$ 
& $P_c^{\Delta}$   & $\frac{3}{2}$ & $ 1$ & $(3,0,2)$ & 
$3835 \pm 54 $ \\
& $P_c^{\Sigma^*}$ & $1$           & $ 0$ & $(2,1,2)$ & 
$4005 \pm 52 $ \\
& $P_c^{\Xi^*}$    & $\frac{1}{2}$ & $-1$ & $(1,2,2)$ & 
$4176 \pm 49 $ \\
& $P_c^{\Omega}$   & $0$           & $-2$ & $(0,3,2)$ & 
$4347 \pm 46 $ \\
$\left[ \phi_{F_2} \times \chi_{F_2} \right]_{F_2}$ 
& $P_c^{\Delta}$   & $\frac{3}{2}$ & $ 1$ & $(3,0,2)$ & 
$3835 \pm 54 $ \\
& $P_c^{\Sigma^*}$ & $1$           & $ 0$ & $(2,1,2)$ & 
$4005 \pm 52 $ \\
& $P_c^{\Xi^*}$    & $\frac{1}{2}$ & $-1$ & $(1,2,2)$ & 
$4176 \pm 49 $ \\
& $P_c^{\Omega}$   & $0$           & $-2$ & $(0,3,2)$ & 
$4347 \pm 46  $ \\
\hline
$\left[ \phi_{F_2} \times \chi_{A_1} \right]_{F_2}$ 
& $P_c^{N}$       & $\frac{1}{2}$ & $ 1$ & $(3,0,2)$ & 
$3882 \pm 57 $ \\
& $P_c^{\Sigma}$  & $1$           & $ 0$ & $(2,1,2)$ & 
$4010 \pm 52$ \\
& $P_c^{\Lambda}$ & $0$           & $ 0$ & $(2,1,2)$ & 
$4037 \pm 54 $ \\
& $P_c^{\Xi}$     & $\frac{1}{2}$ & $-1$ & $(1,2,2)$ & 
$4180\pm  50$ \\
$\left[ \phi_{F_2} \times \chi_{F_2} \right]_{F_2}$ 
& $P_c^{N}$       & $\frac{1}{2}$ & $ 1$ & $(3,0,2)$ & 
$3882 \pm 57 $ \\
& $P_c^{\Sigma}$  & $1$           & $ 0$ & $(2,1,2)$ & 
$4010 \pm 52 $ \\
& $P_c^{\Lambda}$ & $0$           & $ 0$ & $(2,1,2)$ & 
$4037 \pm 54 $ \\
& $P_c^{\Xi}$     & $\frac{1}{2}$ & $-1$ & $(1,2,2)$ & 
$4180 \pm 50   $ \\
$\left[ \phi_{E} \times \chi_{F_2} \right]_{F_2}$ 
& $P_c^{N}$       & $\frac{1}{2}$ & $ 1$ & $(3,0,2)$ & 
$3873 \pm 57 $ \\
& $P_c^{\Sigma}$  & $1$           & $ 0$ & $(2,1,2)$ & 
$4001 \pm 52 $ \\
& $P_c^{\Lambda}$ & $0$           & $ 0$ & $(2,1,2)$ & 
$4027 \pm 54 $ \\
& $P_c^{\Xi}$     & $\frac{1}{2}$ & $-1$ & $(1,2,2)$ & 
$4171 \pm 50 $ \\
$\left[ \phi_{F_1} \times \chi_{F_2} \right]_{F_2}$ 
& $P_c^{N}$       & $\frac{1}{2}$ & $ 1$ & $(3,0,2)$ & 
$3863 \pm  58$ \\
& $P_c^{\Sigma}$  & $1$           & $ 0$ & $(2,1,2)$ & 
$3991\pm  52$ \\
& $P_c^{\Lambda}$ & $0$           & $ 0$ & $(2,1,2)$ & 
$4018 \pm 54$ \\
& $P_c^{\Xi}$     & $\frac{1}{2}$ & $-1$ & $(1,2,2)$ & 
$4162 \pm 50$  
\end{tabular}
\end{ruledtabular}
\end{table}
The results are shown 
in Table~\ref{tab:pentaquarkstates}
following the notation introduced in Fig.~\ref{notation}.
Mean values and uncertainties
are computed through the bootstrap technique
as it was explained in Sec.~\ref{sec:fit}.
Hence, uncertainties in the parameters and their correlations
are carried in full to the computation of the spectrum.
We note that for any given 
$\left[ \phi \times \chi \right]_{F_2}$ state, the 
$P_c^\Lambda$ state has a mean value larger than 
their corresponding $P_c^\Sigma$ partner.
This is at odds with the expectation from the 
$\Lambda-\Sigma$ and $\Lambda_c-\Sigma_c$ mass splittings
(see Table~\ref{tab:groundstates})
where the mass of the $\Sigma$'s are larger.
Although, if we take into account the uncertainties 
we see that hey are equal within errors and
the mass hierarchy and splitting between these 
states cannot be established.
Besides, $\Lambda$ singlets and $\Lambda_c$
are poorly reproduced in our model and the former
were suppressed in the determination
of the parameters.
This impacts the precision of the determination of the
$P_c^\Lambda$ states.
We note that if we directly substitute
the central value parameters
(Table~\ref{tab:parameters}) 
in the pentaquark mass formula,\footnote{We also note that,
statistically speaking, it is not the correct calculation
of the expected value of an observable.}
we obtain that the the hierarchy is inverted with an average mass difference
of $10$~MeV, {\it i.e.} the mass of the $P_c^\Sigma$ is larger than the mass of the $P_c^\Lambda$ for a given $\left[ \phi \times \chi \right]_{F_2}$ state). Nevertheless, the states have equal mass when uncertainties are taking into account. Moreover, central mass values obtained through bootstrap
and through direct substitution of the central values of the parameters are less than $1\sigma$ apart, which indicates a good consistency of the bootstrap method~\cite{EfroTibs93}.

We find that all the ground $P_c^N$ states have lower masses than the LHCb pentaquark candidates ($4380$ and $4450$ MeV), suggesting that these states are excited states. Moreover, because all $P_c^N$ states fall below the $J/\psi\, p$ threshold, their measurement would be very challenging. The rest of the octet and decuplet pentaquarks fall below the thresholds of the obvious two-body channels that could be used for their detection ($\Delta J/\psi$, $\Lambda J/\psi$, $\Sigma J/\psi$, etc.). Hence, if the LHCb pentaquarks are confirmed experimentally in other reactions, the search for other pentaquarks should focus on excited states ($L^\pi \not= 0^+$) instead of on the ground states.

For a discussion of radially excited pentaquark states, in principle, one can use the same mass formula of Eqs.~(\ref{msq})-(\ref{msqorb}). The parameters in the mass formula were fitted to the masses of both ground-state and radially excited $qqq$ baryons. In the same way, the mass formula can be used for radially excited pentaquark states. In that case, there will be a non-vanishing contribution from the orbital part of the mass operator in Eq.~(\ref{msqorb}). Moreover, the excited states carry different quantum numbers as the ground-state pentaquarks which leads to a different contribution from the spin-flavor part. An analysis of excited pentaquark states is in progress. 

A final comment concerns a comparison with the prediction of the masses of hidden-charm pentaquarks of \cite{Yuan:2012wz} which was published before the experimental results of the LHCb Collaboration \cite{Aaij:2016phn,Aaij:2015tga}. The difference between the results for the ground-state hidden-charm pentaquark states in \cite{Yuan:2012wz} and the present study is mainly due to the value of the mass of the charm quark, 
$m_c=1652$ MeV {\it vs.} $1450$ MeV, respectively, which gives rise to a difference in mass of about 400 MeV for the hidden-charm pentaquark states.

\subsection{Magnetic moments}
\label{sec:magneticmoments}
In this section, we study the magnetic moments of the ground state pentaquark states with $J^P=3/2^-$ discussed in the previous section. The magnetic moment is a crucial ingredient in calculation of $J/\psi$ photoproduction cross sections which can provide an independent observation of the pentaquark states \cite{Wang:2015jsa,Kubarovsky:2015aaa,Karliner:2015voa,Wang:2016dzu,Blin:2016dlf,Fernandez-Ramirez:2017gzc,Blin:2018dkm,HallApc,HallBpc,Meziani:2016lhg,HallDpc}.
In Ref.~\cite{Wang:2016dzu}, the magnetic moments were calculated for a molecular model, a diquark-triquark model and a diquark-diquark-antiquark model of pentaquark configurations. To the best of our knowledge, the present calculation is the first one for a CQM of pentaquarks. 

The magnetic moment of a multiquark system is given by the sum of the magnetic moments of its constituent parts~\cite{Bijker:2004gr}
\ba
\vec{\mu} = \vec{\mu}_{\rm spin} + \vec{\mu}_{\rm orb} = \sum_i \mu_i(2\vec{s}_i + \vec{l}_i) ~, 
\label{mmop}
\ea
where $\mu_i$ denotes the magnetic moment of the $i-$th quark or antiquark. Since, at present, we do not consider radial or orbital excitations, the magnetic moment only depends on the spin part. 

The magnetic moments can be obtained from the expectation value of Eq.~\eqref{mmop} using the pentaquark wave function 
of Eq.~\eqref{wfneg} in combination with the spin-flavor wave functions of Appendix~\ref{app:spin-flavor}, the spin wave functions of Appendix~\ref{app:spin}, and
the flavor wave functions of Appendix~\ref{app:flavor1}
\begin{equation}
\mu =
\left< \psi | \,\vec{\mu}\, | \psi \right>
= \frac{1}{3} \sum_{\alpha=\rho,\lambda,\eta} 
\left< \psi^{\rm sf}_{F_{2\alpha}} | \,\vec{\mu}\, | 
\psi^{\rm sf}_{F_{2\alpha}} \right> ~.
\end{equation}

The results for the different $uudc\bar{c}$ configurations are given 
in Table~\ref{magmom}. The numerical values are obtained by using
\begin{align}
\mu_u =& +1.852 \;\mu_N, &
\mu_d =& -0.972 \;\mu_N ~, \nonumber\\
\mu_s =& -0.613 \;\mu_N ~, &
\mu_c =& +0.432 \pm 0.007 \;\mu_N ~. \nonumber
\end{align}
The magnetic moments of the $u$, $d$ and $s$ quarks were fitted to the magnetic moments of the proton, neutron and $\Lambda$ hyperon. The value of the magnetic moment of the charm quark is consistent with the  constituent mass of $m_c=1400 - 1500$ MeV from Table~\ref{tab:parameters}. 
\begin{table}
\caption{Magnetic moments of $uudc\bar{c}$ pentaquark states in $\mu_N$.}
\label{magmom}
\begin{ruledtabular}
\begin{tabular}{cccr}
\noalign{\smallskip}
State & Name & Magnetic moment & Value \\
\noalign{\smallskip}
\hline
\noalign{\smallskip}
$\left[ \phi_{A_1} \times \chi_{F_2} \right]_{F_2}$ & $P_c^{\Delta}$ 
& $\frac{1}{2}(2\mu_u+\mu_d+\mu_c)+\mu_{\bar{c}}$ & 1.164 \\
\noalign{\smallskip}
$\left[ \phi_{F_2} \times \chi_{A_1} \right]_{F_2}$ & $P_c^{\Delta}$ 
& $\frac{9}{10}(2\mu_u+\mu_d+\mu_c)-\frac{3}{5}\mu_{\bar{c}}$ & 3.065 \\
\noalign{\smallskip}
$\left[ \phi_{F_2} \times \chi_{F_2} \right]_{F_2}$ & $P_c^{\Delta}$ 
& $\frac{2}{3}(2\mu_u+\mu_d)+\mu_{\bar{c}}$ & 1.417 \\
\noalign{\smallskip}
\hline
\noalign{\smallskip}
$\left[ \phi_{F_2} \times \chi_{A_1} \right]_{F_2}$ & $P_c^{N}$ 
& $\frac{9}{10}(2\mu_u+\mu_d+\mu_c)-\frac{3}{5}\mu_{\bar{c}}$ & 3.065 \\
\noalign{\smallskip}
$\left[ \phi_{F_2} \times \chi_{F_2} \right]_{F_2}$ & $P_c^{N}$ 
& $\frac{1}{12}(14\mu_u+\mu_d+9\mu_c)+\mu_{\bar{c}}$ & 1.979 \\
\noalign{\smallskip}
$\left[ \phi_{E} \times \chi_{F_2} \right]_{F_2}$ & $P_c^{N}$ 
& $\frac{1}{2}(2\mu_u+\mu_d+\mu_c)+\mu_{\bar{c}}$ & 1.164 \\
\noalign{\smallskip}
$\left[ \phi_{F_1} \times \chi_{F_2} \right]_{F_2}$ & $P_c^{N}$ 
& $\frac{1}{4}(6\mu_u+\mu_d+\mu_c)+\mu_{\bar{c}}$ & 2.232 \\
\noalign{\smallskip}
\end{tabular}
\end{ruledtabular}
\end{table}

\subsection{Electromagnetic couplings}
\label{sec:emcouplings}

For future experiments that aim to study pentaquarks through near threshold $J/\psi$ photoproduction~\cite{HallApc,HallBpc,Meziani:2016lhg,HallDpc}, the size of the electromagnetic couplings of the pentaquarks is important. In this section, we discuss the electromagnetic couplings for the ground state pentaquarks with spin and parity $J^P=3/2^-$.

Electromagnetic couplings are described by 
\begin{equation}
{\cal H}_{\rm em} = e \int d^3 x \hat{J}^{\mu}(\vec{x}) A_{\mu}(\vec{x}) ~,
\label{hem}
\end{equation}
where $J^{\mu}$ is the electromagnetic current summed over all quark flavors 
\begin{equation}
\hat{J}^{\mu}(\vec{x}) = \sum_q e_q \bar{q}(\vec{x}) \gamma^{\mu} q(\vec{x}) ~,
\end{equation}

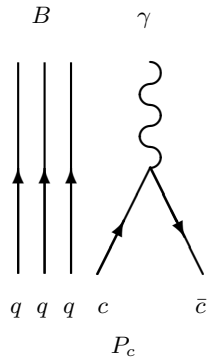
\begin{figure}
\centering
\setlength{\unitlength}{1pt}
\begin{picture}(100,150)(0,0)
\thicklines
\put( 70, 84) {\oval(8,8)[r]}
\put( 70, 92) {\oval(8,8)[l]}
\put( 70,100) {\oval(8,8)[r]}
\put( 70,108) {\oval(8,8)[l]}
\put( 70,116) {\oval(8,8)[r]}
\put( 20,40) {\line(0,1){80}}
\put( 30,40) {\line(0,1){80}}
\put( 40,40) {\line(0,1){80}}
\put( 50,40) {\line( 1,2){20}}
\put( 90,40) {\line(-1,2){20}}
\put( 20,40) {\vector(0,1){40}}
\put( 30,40) {\vector(0,1){40}}
\put( 40,40) {\vector(0,1){40}}
\put( 50,40) {\vector(1, 2){10}}
\put( 70,80) {\vector(1,-2){13}}
\put( 17,25) {$q$}
\put( 27,25) {$q$}
\put( 37,25) {$q$}
\put( 50,25) {$c$}
\put( 87,25) {$\bar{c}$}
\put( 55,10) {$P_c$}
\put( 25,135) {$B$}
\put( 65,135) {$\gamma$}
\end{picture}
\caption{Electromagnetic decay of pentaquark 
$P_c$ into a baryon $B$ and a photon, 
$P_c \rightarrow B + \gamma$.}
\label{emdecay}
\end{figure}

The electromagnetic coupling of Eq.~(\ref{hem}) describes both the emission (and absorption) of the photon off a quark (as is used in studies of photocouplings of baryons, see {\it e.g.} \cite{Copley:1969ft,Close:1979bt,Bijker:2000gq,Bijker:1994yr}), and the annihilation process of a quark-antiquark pair~\cite{LeYaouanc:1988fx}. For the process of interest $P_c \rightarrow B + \gamma$, the relevant term is the annilation of a pair of $c\bar{c}$ quarks. We use the nonrelativistic form of the interaction. The radiative decay widths can be calculated from the helicity amplitudes as \cite{Copley:1969ft,Close:1979bt,Bijker:2000gq} 
\ba
\Gamma(P_c \rightarrow B + \gamma) &=& 
\frac{\rho }{(2\pi)^2} \, \frac{2}{2J+1} \sum_{\nu>0} | A_{\nu}(k) |^2 ~, 
\label{gw}
\ea
where $\rho$ is the phase space and
\ba
A_{\nu}(k) &=& \left< B,\frac{1}{2},\nu-1;\gamma \left| {\cal H}_{\rm em}^{\rm nr} \right| P_c,\frac{3}{2},\nu \right>
\nonumber\\
&=& \sqrt{\frac{4\pi\alpha}{k_0}} \, \beta_{\nu} \, F(k) ~.
\ea
Here $\alpha$ is the fine-structure constant, $k_0$ and $k=|\vec{k}|$ represent the energy and the momentum of the photon. The coefficient $\beta_{\nu}$ is the contribution from the color-spin-flavor part for the annihilation of a $c\bar{c}$ color-singlet pair with spin $S=S_z=1$. Its values are listed in Table~\ref{photo} for different configurations of $uudc\bar{c}$ pentaquarks. It is straightforward to show that for the cases considered, {\it i.e.} ground-state pentaquarks with $J^P=3/2^-$, the helicity amplitudes are related by
\ba
\beta_{1/2} &=& \beta_{3/2}/\sqrt{3} ~.
\ea
Finally, $F(k)$ is a form factor denoting the contribution from the orbital part of the pentaquark wave function. Its form depends on specific type of quark model used: harmonic oscillator, hypercentral, or other. Here we concentrate on the color-spin-flavor part which is common to all quark models. In Table~\ref{photo} we show the results for the contribution from the color-spin-flavor part to the helicity amplitudes. The strongest coupling is to the octet pentaquark configuration with $\phi_{F_1}$, followed by $\phi_{E}$ and $\phi_{F_2}$. The couplings to the octet configuration with $\phi_{A_1}$ and the three decuplet configurations vanish because of symmetry reasons.  

\begin{table}
\caption{Contribution from the color-spin-flavor part to the helicity amplitudes for the electromagnetic decays of $uudc\bar{c}$ pentaquark states into $p + \gamma$ in terms of $e_c=2/3$.}
\label{photo}
\begin{ruledtabular}
\begin{tabular}{cccc}
\noalign{\smallskip}
State & Name & $\beta_{1/2}$ & $\beta_{3/2}$ \\
\noalign{\smallskip}
\hline
\noalign{\smallskip}
$\left[ \phi_{A_1} \times \chi_{F_2} \right]_{F_2}$ & $P_c^{\Delta}$ 
& $0$ & $0$ \\
\noalign{\smallskip}
$\left[ \phi_{F_2} \times \chi_{A_1} \right]_{F_2}$ & $P_c^{\Delta}$ 
& $0$ & $0$ \\
\noalign{\smallskip}
$\left[ \phi_{F_2} \times \chi_{F_2} \right]_{F_2}$ & $P_c^{\Delta}$ 
& $0$ & $0$ \\
\noalign{\smallskip}
\hline
\noalign{\smallskip}
$\left[ \phi_{F_2} \times \chi_{A_1} \right]_{F_2}$ & $P_c^{N}$ 
& $0$ & $0$ \\
\noalign{\smallskip}
$\left[ \phi_{F_2} \times \chi_{F_2} \right]_{F_2}$ & $P_c^{N}$ 
& $\phantom{-}\frac{1}{6\sqrt{2}} e_c$ & $\phantom{-}\frac{1}{2\sqrt{6}} e_c$ \\
\noalign{\smallskip}
$\left[ \phi_{E} \times \chi_{F_2} \right]_{F_2}$ & $P_c^{N}$ 
& $-\frac{1}{6} e_c$ & $-\frac{1}{2\sqrt{3}} e_c$ \\
\noalign{\smallskip}
$\left[ \phi_{F_1} \times \chi_{F_2} \right]_{F_2}$ & $P_c^{N}$ 
& $-\frac{1}{2\sqrt{6}} e_c$ & $-\frac{1}{2\sqrt{2}} e_c$ \\
\noalign{\smallskip}
\end{tabular}
\end{ruledtabular}
\end{table}

\section{Summary and conclusions}
\label{sec:conclusions}

Motivated by the hidden-charm pentaquark candidates recently found by the LHCb Collaboration~\cite{Aaij:2015tga,Aaij:2016phn,Jurik:2016bdm}, 
we presented an analysis of ground state pentaquarks in a constituent quark model. In doing so we first built an $SU(4)$ quark model with a mass operator that is a generalization of the G\"ursey-Radicati mass formula commonly used to study baryon resonances in which the $SU(4)$ flavor symmetry is broken dynamically by the mass difference between the charm quark $c$ and the light quarks, $u$, $d$ and $s$. The parameters, {\it i.e.} constituent quark masses and couplings, were determined by fitting the 16 ground state baryons consisting of $u$, $d$, $s$ and $c$ quarks and 37 $N$, $\Delta$, $\Lambda$, $\Sigma$ and $\Xi$ excitations, all of them with a three- or four-star status in the PDG~\cite{Tanabashi:2018oca}. 
Uncertainties in are computed using the bootstrap technique
that allows to propagate statistical errors from data to parameters
and observables accounting in full for the correlations~\cite{recipes,EfroTibs93,Landay:2016cjw}.
The general structure of the spectrum is well reproduced, with a root mean square deviation of $\delta_{rms}=44 \, \text{MeV}$.
Next we used the same mass formula to predict the double- and triple-charm baryons, obtaining a good agreement with both 
the latest measurement of the $\Xi_{cc}$ state by LHCb~\cite{Aaij:2017ueg} and LQCD 
calculations~\cite{Liu:2009jc,Briceno:2012wt,Namekawa:2013vu,Alexandrou:2014sha,Brown:2014ena,Bali:2015lka,Alexandrou:2017xwd}.

In order to apply the mass formula to pentaquarks we presented a classification of a complete set of basis states for all ground-state hidden-charm pentaquark configurations of the form $qqqc\bar{c}$ with spin and parity $J^P=3/2^-$, where $q$ denotes the light quarks $q=u$, $d$ and/or $s$. As a result it was found that for these $qqqc\bar{c}$ states there are three allowed configurations associated with $SU(3)$ flavor decuplets and four with flavor octets which differ from each other in their flavor and spin content. 

The masses of the ground-state hidden-charm pentaquarks $P_c^N$ with $J^P=3/2^-$ are calculated with a theoretical uncertainty of $50-60$ MeV. The calculated masses are substantially lower than 
those measured by LHCb suggesting that, if indeed the detected signals correspond to $uudc\bar{c}$ pentaquark states, they do not belong to ground state pentaquarks but rather to radially excited configurations. All calculated ground-state octet and decuplet pentaquarks fall below the thresholds of the obvious two-body channels that could be used for their detection
($\Delta J/\psi$, $\Lambda J/\psi$, $\Sigma J/\psi$, etc.).

Finally, we computed magnetic moments and photocouplings of the ground-state pentaquarks. These quantities are of interest to estimate the pentaquark photoproduction cross
section~\cite{Kubarovsky:2015aaa,Karliner:2015voa,Wang:2016dzu,Blin:2016dlf,Fernandez-Ramirez:2017gzc,Blin:2018dkm} which are relevant to the approved experiments at Jefferson Lab on $J/\psi$ photoproduction~\cite{HallApc,HallBpc,Meziani:2016lhg,HallDpc}
to try to find the $P_c^N(4450)$ signal with an electromagnetic probe. We note that because of the symmetry of the pentaquark wave function several configurations cannot be excited in photoproduction, such as the $P_c^\Delta$ configurations belonging to the $SU(3)$ decuplet. For the ground state pentaquarks, there are three configurations with nonvanishing photocouplings. An analysis of the electromagnetic couplings of excited pentaquark states is in progress. 

In conclusion, we presented a detailed analysis of ground-state pentaquark states in the context of the quark model, and identified several configurations that, in principle, can be excited in future photoproduction experiments. 

\begin{acknowledgments}
This work was supported in part by PAPIIT-DGAPA (UNAM, Mexico) grants No.~IN109017,  No.~IA101717, and No.~IA101819, and by CONACYT (Mexico) grants No.~251817 and No.~340629.
\end{acknowledgments}

\appendix
\section{Generalized Gell-Mann-Okubo formula}\label{app:GMO}
The masses of the light quarks $u$, $d$ and $s$ are 
much smaller than that of the charm quark $c$. 
Let's consider the case in which
\ba
m_u = m_d = m_s = m < m_c ~.
\ea
In this case, the expectation value of 
$H_{\rm strong}$ in flavor space is given 
by the quark mass matrix of the form  
\ba
\left< H_{\rm strong} \right> 
&=& \left( \begin{array}{cccc} m & 0 & 0 & 0 \\ 0 & m & 0 & 0 \\ 
0 & 0 & m & 0 \\ 0 & 0 & 0 & m_c \end{array} \right) 
\nonumber\\
&=& \frac{3m+m_c}{4} 
\left( \begin{array}{cccc} 1 & 0 & 0 & 0 \\ 0 & 1 & 0 & 0 \\ 
0 & 0 & 1 & 0 \\ 0 & 0 & 0 & 1 \end{array} \right) 
\nonumber\\
&& + \frac{m-m_c}{4} \left( 
\begin{array}{cccc} 1 & 0 & 0 & 0 \\ 0 & 1 & 0 & 0 \\ 0 & 0 & 1 & 0 \\ 
0 & 0 & 0 & -3 \end{array} \right) 
\nonumber\\
&=& \frac{3m+m_c}{4} {\bf 1} + \frac{m-m_c}{4} \lambda_{15},
\label{qmass}
\ea
where $\lambda_{15}$ denotes one of the 
Gell-Mann matrices for $SU(4)$ 
satisfying the Lie algebra \cite{Stancu:1991rc} 
\ba
\, [\lambda_a,\lambda_b] &=& 2i \, f_{abc} \lambda_c ~.  
\ea 
The first term in Eq.~\eqref{qmass} is an invariant 
under the $SU_{\rm f}(4)$ flavor group, 
whereas the second term transforms like $\lambda_{15}$, 
{\it i.e.} the 15th component of a 
15-dimensional multiplet. In general, 
one can express $\left< H_{\rm strong} \right>$ as
\ba
\left< H_{\rm strong} \right> &=& \left< H_0 + H_{15} \right> ~,
\ea
where $H_0$ is a scalar under $SU_{\rm f}(4)$ 
and $H_{15}$ transforms as $\lambda_{15}$. 
The most general form of $H_{15}$ is 
\ba
H_{15} &=& x \, \hat{F}_{15} + y \, d_{15,ab} \hat{F}_a \hat{F}_b ~,
\label{h15}
\ea
where $x$ and $y$ are coefficients, 
$\hat{F}_a=\lambda_a/2$ are the $SU_{\rm f}(4)$ generators, 
and the coefficients $d_{abc}$ are defined by
\ba
\{\lambda_a,\lambda_b\} &=& \delta_{ab} {\bf 1} + 2d_{abc} 
\lambda_c ~. 
\ea
Equation~\eqref{h15} can be expressed in terms 
of $\hat{Z}=\sqrt{3/2}\, \hat{F}_{15}$ and the Casimir 
invariants of the flavor groups as  
\begin{equation}
\begin{split}
H_{15} =& x \sqrt{\frac{2}{3}} \hat{Z} 
+ y \frac{1}{\sqrt{6}} \left[ 
\left( \sum_{i=1}^8 - \sum_{i=9}^{14} \right) \hat{F}_i^2 
- 2 \hat{F}_{15}^2 \right] \\
=& x \sqrt{\frac{2}{3}} \hat{Z} - y \frac{1}{\sqrt{6}} 
\left[ \left( \sum_{i=1}^{15} - 2 \sum_{i=1}^{8} \right) 
\hat{F}_i^2 + \hat{F}_{15}^2 \right] \\
=& x \sqrt{\frac{2}{3}} \hat{Z} - y \frac{1}{\sqrt{6}} 
\left( \hat{C}_{2SU_{\rm f}(4)} 
- 2 \hat{C}_{2SU_{\rm f}(3)} + \frac{2}{3} \hat{Z}^2 \right) ~.
\end{split}
\end{equation}
with eigenvalues 
\begin{equation}
\begin{split}
\left< H_{15} \right> = &\, m_0 + \delta m_1 \, Z \\ 
+ &\, \delta m_2 \, \frac{1}{3} \left[ \lambda(\lambda+3) 
+ \mu(\mu+3) + \lambda \mu - Z^2 \right] ~, 
\end{split}\label{GMO4}
\end{equation}
with $\delta m_1 = x \sqrt{2/3}$ and $\delta m_2 = y \sqrt{2/3}$. 
The expectation value of $C_{2SU_{\rm f}(4)}$ 
for a given $SU_{\rm f}(4)$ multiplet $[g]$ 
is absorbed into $m_0$. 

This form is very similar to the Gell-Mann-Okubo mass formula 
for three flavors (in the notation of~\cite{nachtmann1990elementary})
\begin{equation}
\left< H_{8} \right> = m'_0 + \delta m'_1 \, Y  
+ \delta m'_2 \, \left[ I(I+1) - \frac{1}{4} Y^2 \right] ~. 
\label{GMO3}
\end{equation}

\section{Spin wave functions} 
\label{app:spin}
The spin wave function of quark is given 
by the Young tableau $[1]$ which corresponds to 
$S=1/2$ with projection $M_S=1/2$ ($\uparrow$) 
or $-1/2$ ($\downarrow$). 
The spin wave functions for pentaquarks with spin 
$S^{\cal P}=3/2^-$ can be obtained in a systematic 
way by using the Clebsch-Gordan series 
with the Condon-Shortley phase convention. 
The results are given in Table~\ref{spinwf}. 
The subindices 
refer to the symmetry properties of the four-quark 
subsystem under permutation. 
\begin{table}
\caption[]{Pentaquark spin wave functions 
for states with $S^{\cal P}=3/2^-$ and $M=3/2$.}
\label{spinwf}
\begin{ruledtabular}
\begin{tabular}{lcl}
$\left| \chi_{A_1} \right>$ &=& $\left| \frac{1}{2\sqrt{5}} 
\left( 4 \uparrow \uparrow \uparrow \uparrow \downarrow 
       - \left( \uparrow \uparrow \uparrow \downarrow 
              + \uparrow \uparrow \downarrow \uparrow  
              + \uparrow \downarrow \uparrow \uparrow  
              + \downarrow \uparrow \uparrow \uparrow \right) \uparrow \right) \right>$ \\
$\left| \chi_{F_{2\rho}} \right>$ &=& $\left| \frac{1}{\sqrt{2}} 
\left( \uparrow \downarrow \uparrow \uparrow 
     - \downarrow \uparrow \uparrow \uparrow \right) \uparrow \right>$ \\
$\left| \chi_{F_{2\lambda}} \right>$ &=& $\left| \frac{1}{\sqrt{6}} 
\left( 2 \uparrow \uparrow \downarrow \uparrow 
       - \uparrow \downarrow \uparrow \uparrow   
       - \downarrow \uparrow \uparrow \uparrow \right) \uparrow \right>$ \\
$\left| \chi_{F_{2\eta}} \right>$ &=& $\left| \frac{1}{2\sqrt{3}} 
\left( 3 \uparrow \uparrow \uparrow \downarrow 
       - \uparrow \uparrow \downarrow \uparrow  
       - \uparrow \downarrow \uparrow \uparrow  
       - \downarrow \uparrow \uparrow \uparrow \right) \uparrow \right>$
\end{tabular}
\end{ruledtabular}
\end{table}
%
\section{Flavor wave functions}
In this appendix we discuss the flavor wave functions for 
the two coupling schemes discussed in Sec.~\ref{sec:pentaquarks},
corresponding to couplings at the level of the 
three-flavor $SU(3)$ symmetry and the four-flavor 
$SU(4)$ symmetry.
\subsection{Three-flavor SU(3) couplings}
\label{app:flavor1}
The flavor wave functions of the pentaquarks 
of Eqs.~\eqref{penta3} can be constructed by acting 
with the ladder operators in flavor space 
on the highest weight state. 

As an example, consider the four-quark system 
consisting of four up quarks which has isospin 
$I=I_3=2$ and hypercharges, $Y=4/3$ and $Z=1$. 
This state is obviously symmetric under 
any permutation: $[g]=[4]$ and $[h]=[4]$. 
In explicit form, the flavor wave function 
$|[g],[h],I,I_3,Y,Z \rangle$ is given by 
\ba
\left| [4],[4],2,2,\frac{4}{3},1 \right>_{A_1} &=& \left| uuuu \right> ~.
\ea
The states with other values of $[h]$, $I$, $I_3$, $Y$ and $Z$ can be obtained by applying the appropriate ladder operators in flavor space. For example, the symmetric $uudc$ state is obtained as follows
\begin{equation}
\begin{split}
&\left| [4],[3],\frac{3}{2},\frac{1}{2},1,0 \right>_{A_1} \\
&\hspace{1cm}= \frac{1}{2\sqrt{3}} \, I_- M_- V_- 
\left| [4],[4],2,2,\frac{4}{3},1 \right>_{A_1}\\
&\hspace{1cm}= \frac{1}{\sqrt{3}} \, I_- M_- \left| [4],[4],
\frac{3}{2},\frac{3}{2},\frac{1}{3},1 \right>_{A_1}\\
&\hspace{1cm}= \frac{1}{\sqrt{3}} \, I_- \left| [4],[3],\frac{3}{2},\frac{3}{2},1,0 \right>_{A_1} \\
&\hspace{1cm}= \Big| \frac{1}{2\sqrt{3}} 
\big( uudc+uduc+duuc+uucd \\
&\hspace{2.1cm} +udcu+ducu+ucud+ucdu \\
&\hspace{2.1cm} +dcuu+cuud+cudu+cduu \big) \Big> ~.
\label{wfa1}
\end{split}
\end{equation}
Here we have used the phase convention of Baird and Biedenharn~\cite{Baird:1963wv}, in combination with the notation of the lowering operators of Haacke {\it et al.}~\cite{Haacke:1975rt}
\ba
I_- |u\rangle &=& |d\rangle ~,
\nonumber\\
U_- |d\rangle &=& |s\rangle ~,
\nonumber\\
M_- |s\rangle &=& |c\rangle ~.
\ea
Finally, the symmetric pentaquark wave functions of Eqs.~\eqref{wf3} and~\eqref{penta3} are obtained by coupling Eq.~\eqref{wfa1} to the wave function of the antiquark with charm of Eq.~\eqref{antiquarkwf}. 

The other pentaquark flavor wave functions corresponding to the $q^4$ configurations of Eqs.~\eqref{qqqqflavor10} and \eqref{qqqqflavor8} can be obtained in a similar way. 
Table~\ref{flavorwf} shows the results for all possible 12 multiquark $uudc\bar{c}$ configurations. 
The results for the octet configurations coincide with the product of the four-quark $uudc$ wave functions of Appendix~A.2~\cite{Yuan:2012wz} and that of the $\bar{c}$ quark which in the Baird and Biedenharn phase convention carries a minus sign, see Eq.~(\ref{antiquarkwf}).  

\begin{table*}
\caption[]{Flavor wave functions of $uudc\bar{c}$ 
pentaquarks for the decuplet (top) and the octet (bottom).}
\label{flavorwf}
\begin{ruledtabular}
\begin{tabular}{lcl}
\noalign{\smallskip}
$\left| \phi_{A_1} \right>$ &=& $\left| -\frac{1}{2\sqrt{3}}
\left(uudc+uduc+duuc+uucd+udcu+ducu+ucud+ucdu+dcuu+cuud+cudu+cduu \right) \bar{c} \right>$ \\ 
\noalign{\smallskip}
$\left| \phi_{F_{2\rho}} \right>$ &=& $\left| -\frac{1}{\sqrt{6}} \left( dcuu+ucdu+ucud-cduu-cudu-cuud \right) \bar{c} \right>$ \\
\noalign{\smallskip}
$\left| \phi_{F_{2\lambda}} \right>$ &=& $\left| -\frac{1}{3\sqrt{2}} \left( 2ducu+2udcu+2uucd-dcuu-ucdu-ucud-cduu-cudu-cuud \right) \bar{c} \right>$ \\
\noalign{\smallskip}
$\left| \phi_{F_{2\eta}} \right>$ &=& $\left| -\frac{1}{6} \left( 3duuc+3uduc+3uudc-ducu-udcu-uucd-dcuu-ucdu-ucud-cduu-cudu-cuud \right) \bar{c} \right>$ \\
\noalign{\smallskip}
\hline
\noalign{\smallskip}
$\left| \phi_{F_{2\rho}} \right>$ &=& $\left| -\frac{1}{4\sqrt{3}} \left( 3(udcu+uduc-ducu-duuc)+2(cduu-dcuu)+ucdu+ucud-cudu-cuud \right) \bar{c} \right>$ \\
\noalign{\smallskip}
$\left| \phi_{F_{2\lambda}} \right>$ &=& $\left| -\frac{1}{12} \left( 6uudc+2uucd+5(cudu+ucdu)-4(dcuu+cduu)-3(duuc+uduc)-ducu-udcu-cuud-ucud \right) \bar{c} \right>$ \\
\noalign{\smallskip}
$\left| \phi_{F_{2\eta}} \right>$ &=& $\left| -\frac{1}{3\sqrt{2}} \left( 2(cuud+ucud+uucd)-dcuu-ducu-cduu-udcu-cudu-ucdu  \right) \bar{c} \right>$ \\
\noalign{\smallskip}
$\left| \phi_{E_{\rho}} \right>$ &=& $\left| -\frac{1}{2\sqrt{2}} \left( ucud+uduc-cuud-duuc-ucdu-udcu+cudu+ducu \right) \bar{c} \right>$ \\
\noalign{\smallskip}
$\left| \phi_{E_{\lambda}} \right>$ &=& $\left| -\frac{1}{2\sqrt{6}} \left( 2uucd+2uudc-ucud-uduc-cuud-duuc+2cduu+2dcuu-ucdu-udcu-cudu-ducu \right) \bar{c} \right>$ \\
\noalign{\smallskip}
$\left| \phi_{F_{1\rho}} \right>$ &=& $\left| -\frac{1}{4\sqrt{3}} \left( 2dcuu-2cduu-udcu+ducu+ucdu-cudu+3uduc-3duuc-3ucud+3cuud \right) \bar{c} \right>$ \\
\noalign{\smallskip}
$\left| \phi_{F_{1\lambda}} \right>$ &=& $\left| -\frac{1}{4} \left( 2uudc-uduc-duuc-2uucd+ucud+cuud+udcu+ducu-ucdu-cudu \right) \bar{c} \right>$ \\
\noalign{\smallskip}
$\left| \phi_{F_{1\eta}} \right>$ &=& $\left| -\frac{1}{\sqrt{6}} \left( dcuu-cduu-ducu+udcu+cudu-ucdu \right) \bar{c} \right>$ \\
\noalign{\smallskip}
\end{tabular}
\end{ruledtabular}
\end{table*}
\subsection{Four-flavor SU(4) couplings}
\label{app:flavor2}
The flavor wave functions of Eqs.~\eqref{decuplet4} 
and \eqref{octet4} can be expanded into a basis of 
product states of four-quark states and one antiquark 
state by inserting the appropriate $SU(4)$ 
Clebsch-Gordan coefficients~\cite{Alex:2010wi}. 
The pentaquark states with $A_1$ symmetry were already 
discussed in Sec.~\ref{sec:PentaquarkStates}.

For the pentaquark states with $F_2$ symmetry we find
\ba
&& \left| [421],[411],\frac{3}{2},\frac{1}{2},1,\frac{3}{4} \right>_{F_2} 
\nonumber\\ 
&& \hspace{1cm} = 
-\sqrt{\frac{2}{3}} \left| [31],[3],\frac{3}{2},\frac{1}{2},1,0 \right>_{F_2} \left| \phi(\bar{c}) \right>  
\nonumber\\
&& \hspace{1.2cm} +\frac{1}{\sqrt{6}} \left| [31],[31],\frac{3}{2},\frac{1}{2},\frac{1}{3},1 \right>_{F_2} \left| \phi(\bar{s}) \right>  
\nonumber\\
&& \hspace{1.2cm} +\frac{1}{3} \left| [31],[31],1,0,\frac{4}{3},1 \right>_{F_2} \left| \phi(\bar{d}) \right>  
\nonumber\\
&& \hspace{1.2cm} -\frac{1}{3\sqrt{2}} \left| [31],[31],1,1,\frac{4}{3},1 \right>_{F_2} \left| \phi(\bar{u}) \right> ~, \label{F2d1}
\ea
\ba
&& \left| [4111],[411],\frac{3}{2},\frac{1}{2},1,\frac{3}{4} \right>_{F_2} \nonumber\\
&& \hspace{1cm} = 
-\frac{1}{\sqrt{3}} \left| [31],[3],\frac{3}{2},\frac{1}{2},1,0 \right>_{F_2} \left| \phi(\bar{c}) \right>  
\nonumber\\
&& \hspace{1.2cm} -\frac{1}{\sqrt{3}} \left| [31],[31],\frac{3}{2},\frac{1}{2},\frac{1}{3},1 \right>_{F_2} \left| \phi(\bar{s}) \right>  
\nonumber\\
&& \hspace{1.2cm} -\frac{\sqrt{2}}{3} \left| [31],[31],1,0,\frac{4}{3},1 \right>_{F_2} \left| \phi(\bar{d}) \right>  
\nonumber\\
&& \hspace{1.2cm} +\frac{1}{3} \left| [31],[31],1,1,\frac{4}{3},1 \right>_{F_2} \left| \phi(\bar{u}) \right> ~, \label{F2d2} 
\ea
and
\ba
&& \left| [421],[321],\frac{1}{2},\frac{1}{2},1,\frac{3}{4} \right>_{F_2} 
\nonumber\\
&& \hspace{1cm} = 
-\sqrt{\frac{5}{6}} \left| [31],[21],\frac{1}{2},\frac{1}{2},1,0 \right>_{F_2} \left| \phi(\bar{c}) \right>  
\nonumber\\
&& \hspace{1.2cm} +\frac{1}{\sqrt{30}} \left| [31],[31],\frac{1}{2},\frac{1}{2},\frac{1}{3},1 \right>_{F_2} \left| \phi(\bar{s}) \right>  
\nonumber\\
&& \hspace{1.2cm} +\frac{\sqrt{2}}{3\sqrt{5}} \left| [31],[31],1,0,\frac{4}{3},1 \right>_{F_2} \left| \phi(\bar{d}) \right>  
\nonumber\\
&& \hspace{1.2cm} +\frac{2}{3\sqrt{5}} \left| [31],[31],1,1,\frac{4}{3},1 \right>_{F_2} \left| \phi(\bar{u}) \right> ~, \label{F2o1}  
\ea
\ba
&& \left| [3211],[321],\frac{1}{2},\frac{1}{2},1,\frac{3}{4} \right>_{F_2} 
\nonumber\\
&& \hspace{1cm} = 
\frac{1}{\sqrt{6}} \left| [31],[21],\frac{1}{2},\frac{1}{2},1,0 \right>_{F_2} \left| \phi(\bar{c}) \right>  
\nonumber\\
&& \hspace{1.2cm} +\frac{1}{\sqrt{6}} \left| [31],[31],\frac{1}{2},\frac{1}{2},\frac{1}{3},1 \right>_{F_2} \left| \phi(\bar{s}) \right>  
\nonumber\\
&& \hspace{1.2cm} +\frac{\sqrt{2}}{3} \left| [31],[31],1,0,\frac{4}{3},1 \right>_{F_2} \left| \phi(\bar{d}) \right>  
\nonumber\\
&& \hspace{1.2cm} +\frac{2}{3} \left| [31],[31],1,1,\frac{4}{3},1 \right>_{F_2} \left| \phi(\bar{u}) \right> ~, \label{F2o2} 
\ea
which can be combined to give the $uudc\bar{c}$ configurations with $F_2$ symmetry of Eqs.~(\ref{wfF2d}) and (\ref{wfF2o}). 

For the pentaquark states with $E$ symmetry one has
\ba
&& \left| [331],[321],\frac{1}{2},\frac{1}{2},1,\frac{3}{4} \right>_{E}
\nonumber\\
&& \hspace{1cm} = 
-\frac{\sqrt{3}}{2} \left| [22],[21],\frac{1}{2},\frac{1}{2},1,0 \right>_{E} \left| \phi(\bar{c}) \right>  
\nonumber\\
&& \hspace{1.2cm} +\frac{1}{2\sqrt{3}} \left| [22],[22],\frac{1}{2},\frac{1}{2},\frac{1}{3},1 \right>_{E} \left| \phi(\bar{s}) \right> 
\nonumber\\
&& \hspace{1.2cm} +\frac{1}{\sqrt{6}} \left| [22],[22],0,0,\frac{4}{3},1 \right>_{E} \left| \phi(\bar{d}) \right> ~, \label{Eo1}
\ea
\ba
&& \left| [3211],[321],\frac{1}{2},\frac{1}{2},1,\frac{3}{4} \right>_{E} 
\nonumber\\
&& \hspace{1cm} = 
-\frac{1}{2} \left| [22],[21],\frac{1}{2},\frac{1}{2},1,0 \right>_{E} \left| \phi(\bar{c}) \right>  
\nonumber\\
&& \hspace{1.2cm} -\frac{1}{2} \left| [22],[22],\frac{1}{2},\frac{1}{2},\frac{1}{3},1 \right>_{E} \left| \phi(\bar{s}) \right>  
\nonumber\\
&& \hspace{1.2cm} -\frac{1}{\sqrt{2}} \left| [22],[22],0,0,\frac{4}{3},1 \right>_{E} \left| \phi(\bar{d}) \right> ~, \label{Eo2}
\ea
which can be combined to give the $uudc\bar{c}$ configuration with $E$ symmetry of Eq.~\eqref{wfEo}. 

Finally, for the pentaquark states with $F_1$ symmetry one finds
\ba
&& \left| [322],[321],\frac{1}{2},\frac{1}{2},1,\frac{3}{4} \right>_{F_1} 
\nonumber\\
&& \hspace{1cm} = 
-\frac{1}{\sqrt{2}} \left| [211],[21],\frac{1}{2},\frac{1}{2},1,0 \right>_{F_1} \left| \phi(\bar{c}) \right>  
\nonumber\\
&& \hspace{1.2cm} +\frac{1}{\sqrt{2}} \left| [211],[211],\frac{1}{2},\frac{1}{2},\frac{1}{3},1 \right>_{F_1} \left| \phi(\bar{s}) \right> ~, \label{F1o1}  
\ea
\ba
&& \left| [3211],[321],\frac{1}{2},\frac{1}{2},1,\frac{3}{4} \right>_{F_1} 
\nonumber\\
&& \hspace{1cm} = 
+\frac{1}{\sqrt{2}} \left| [211],[21],\frac{1}{2},\frac{1}{2},1,0 \right>_{F_1} \left| \phi(\bar{c}) \right>  
\nonumber\\
&& \hspace{1.2cm} +\frac{1}{\sqrt{2}} \left| [211],[211],\frac{1}{2},\frac{1}{2},\frac{1}{3},1 \right>_{F_1} \left| \phi(\bar{s}) \right> ~, \label{F1o2}
\ea
which can be combined to give the $uudc\bar{c}$ configuration with $F_1$ symmetry of Eq.~\eqref{wfF1o}.

\section{Spin-flavor wave functions}
\label{app:spin-flavor}
The explicit expressions of the spin-flavor wave functions 
in terms of the spin and flavor 
wave functions of Tables~\ref{spinwf} and~\ref{flavorwf} 
are given by using the isoscalar factors for the decomposition 
${\cal T}_d \sim S_4 \supset S_3 \supset S_2$~\cite{Stancu:1991rc}. 
As a result, we find 
\ba
\psi^{\rm sf}_{F_{2\alpha}} &=& \left[ \phi_{A_1} \times \chi_{F_2} \right]_{F_{2\alpha}} \;=\; \phi_{A_1} \chi_{F_{2\alpha}} ~, 
\nonumber\\
\psi^{\rm sf}_{F_{2\alpha}} &=& \left[ \phi_{F_2} \times \chi_{A_1} \right]_{F_{2\alpha}} \;=\; \phi_{F_{2\alpha}} \chi_{A_1} ~,
\ea
with $\alpha=\rho$, $\lambda$, $\eta$, and 
\ba 
\psi^{\rm sf}_{F_{2\rho}} &=& \left[ \phi_{F_2} \times \chi_{F_2} \right]_{F_{2\rho}} 
\nonumber\\
&=& -\frac{1}{\sqrt{3}} \left[ \phi_{F_{2\rho}} \chi_{F_{2\lambda}} 
+ \phi_{F_{2\lambda}} \chi_{F_{2\rho}} \right] 
\nonumber\\
&& -\frac{1}{\sqrt{6}} \left[ \phi_{F_{2\rho}} \chi_{F_{2\eta}} 
+ \phi_{F_{2\eta}} \chi_{F_{2\rho}} \right] ~, 
\nonumber\\
\psi^{\rm sf}_{F_{2\lambda}} &=& \left[ \phi_{F_2} \times \chi_{F_2} \right]_{F_{2\lambda}} 
\nonumber\\
&=& -\frac{1}{\sqrt{3}} \left[ \phi_{F_{2\rho}} \chi_{F_{2\rho}} 
- \phi_{F_{2\lambda}} \chi_{F_{2\lambda}} \right] 
\nonumber\\
&& -\frac{1}{\sqrt{6}} \left[ \phi_{F_{2\lambda}} \chi_{F_{2\eta}} 
+ \phi_{F_{2\eta}} \chi_{F_{2\lambda}} \right] ~, 
\nonumber\\
\psi^{\rm sf}_{F_{2\eta}} &=& \left[ \phi_{F_2} \times \chi_{F_2} \right]_{F_{2\eta}} 
\\
&=& -\frac{1}{\sqrt{6}} \left[ \phi_{F_{2\rho}} \chi_{F_{2\rho}} 
+\phi_{F_{2\lambda}} \chi_{F_{2\lambda}} -2 \phi_{F_{2\eta}} \chi_{F_{2\eta}} \right] ~, \nonumber
\ea
followed by
\ba
\psi^{\rm sf}_{F_{2\rho}} &=& \left[ \phi_{E} \times \chi_{F_2} \right]_{F_{2\rho}} 
\nonumber\\ 
&=& -\frac{1}{2} \left[ \phi_{E_{\rho}} \chi_{F_{2\lambda}} 
+ \phi_{E_{\lambda}} \chi_{F_{2\rho}} \right]
+\frac{1}{\sqrt{2}} \phi_{E_{\rho}} \chi_{F_{2\eta}} ~,
\nonumber\\
\psi^{\rm sf}_{F_{2\lambda}} &=& \left[ \phi_{E} \times \chi_{F_2} \right]_{F_{2\lambda}} 
\nonumber\\ 
&=& -\frac{1}{2} \left[ \phi_{E_{\rho}} \chi_{F_{2\rho}} 
- \phi_{E_{\lambda}} \chi_{F_{2\lambda}} \right]  
+\frac{1}{\sqrt{2}} \phi_{E_{\lambda}} \chi_{F_{2\eta}} ~,
\nonumber\\
\psi^{\rm sf}_{F_{2\eta}} &=& \left[ \phi_{E} \times \chi_{F_2} \right]_{F_{2\eta}} 
\nonumber\\ 
&=& \frac{1}{\sqrt{2}} \left[ \phi_{E_{\rho}} \chi_{F_{2\rho}} 
+ \phi_{E_{\lambda}} \chi_{F_{2\lambda}} \right] ~, 
\label{wfsf}
\ea 
and
\ba 
\psi^{\rm sf}_{F_{2\rho}} &=& \left[ \phi_{F_1} \times \chi_{F_2} \right]_{F_{2\rho}} 
\nonumber\\
&=& -\frac{1}{\sqrt{2}} \left[ \phi_{F_{1\rho}} \chi_{F_{2\eta}} 
+ \phi_{F_{1\eta}} \chi_{F_{2\lambda}} \right] ~, 
\nonumber\\
\psi^{\rm sf}_{F_{2\lambda}} &=& \left[ \phi_{F_1} \times \chi_{F_2} \right]_{F_{2\lambda}} 
\nonumber\\
&=& -\frac{1}{\sqrt{2}} \left[ \phi_{F_{1\lambda}} \chi_{F_{2\eta}} 
- \phi_{F_{1\eta}} \chi_{F_{2\rho}} \right] ~, 
\nonumber\\
\psi^{\rm sf}_{F_{2\eta}} &=& \left[ \phi_{F_1} \times \chi_{F_2} \right]_{F_{2\eta}} 
\nonumber\\
&=& \frac{1}{\sqrt{2}} \left[ \phi_{F_{1\rho}} \chi_{F_{2\rho}} 
+\phi_{F_{1\lambda}} \chi_{F_{2\lambda}} \right] ~. 
\ea

\section{Color wave functions}

The color part of the pentaquark wave functions can be obtained by combining the color wave function of the four-quark system with that of the antiquark
\ba
\psi_{F_{1\alpha}}^{\rm c} &=& \frac{1}{\sqrt{3}} \left[ 
\left| [211],\frac{1}{2},\frac{1}{2},\frac{1}{3} \right>_{\alpha} \bar{r}\right. \nonumber \\
&& +\left| [211],\frac{1}{2},-\frac{1}{2},\frac{1}{3} \right>_{\alpha} \bar{g} \nonumber \\
&& +\left.\left| [211],0,0,-\frac{2}{3} \right>_{\alpha} \bar{b}
\right]\, ,
\ea
with $\alpha=\rho$, $\lambda$, $\eta$.
The color wave functions of $q^4$ system are given in Table~\ref{colorwf}.
\begin{table*}
\caption[]{Color wave functions of $q^4$ configurations with $[211]$ and $F_1$ symmetry.}
\label{colorwf}
\begin{ruledtabular}
\begin{tabular}{lcl}
\noalign{\smallskip}
$\left| [211],\frac{1}{2},\frac{1}{2},\frac{1}{3} \right>_{\rho}$ &=& 
$\left| \frac{1}{4\sqrt{3}} \left[ 
(2gbr-2bgr-rgb+grb+rbg-brg)r+3(rgr-grr)b-3(rbr-brr)g \right] \right>$ \\
\noalign{\smallskip}
$\left| [211],\frac{1}{2},-\frac{1}{2},\frac{1}{3} \right>_{\rho}$ &=& 
$\left| \frac{1}{4\sqrt{3}} \left[ 
(2brg-2rbg-gbr+bgr+grb-rgb)g+3(gbg-bgg)r-3(grg-rgg)b \right] \right>$ \\
\noalign{\smallskip}
$\left| [211],0,0,-\frac{2}{3} \right>_{\rho}$ &=& 
$\left| \frac{1}{4\sqrt{3}} \left[ 
(2rgb-2grb-brg+rbg+bgr-gbr)b+3(brb-rbb)g-3(bgb-gbb)r \right] \right>$ \\
\noalign{\smallskip}
\hline
\noalign{\smallskip}
$\left| [211],\frac{1}{2},\frac{1}{2},\frac{1}{3} \right>_{\lambda}$ &=& 
$\left| \frac{1}{4} \left[
(2rrg-rgr-grr)b-(2rrb-rbr-brr)g+(rgb+grb-rbg-brg)r \right] \right>$ \\
\noalign{\smallskip}
$\left| [211],\frac{1}{2},-\frac{1}{2},\frac{1}{3} \right>_{\lambda}$ &=& 
$\left| \frac{1}{4} \left[
(2ggb-gbg-bgg)r-(2ggr-grg-rgg)b+(gbr+bgr-grb-rgb)g \right] \right>$ \\
\noalign{\smallskip}
$\left| [211],0,0,-\frac{2}{3} \right>_{\lambda}$ &=& 
$\left| \frac{1}{4} \left[
(2bbr-brb-rbb)g-(2bbg-bgb-gbb)r+(brg+rbg-bgr-gbr)b \right] \right>$ \\
\noalign{\smallskip}
\hline
\noalign{\smallskip}
$\left| [211],\frac{1}{2},\frac{1}{2},\frac{1}{3} \right>_{\eta}$ &=& 
$\left| \frac{1}{\sqrt{6}} (rgb-grb+brg-rbg+gbr-bgr)r \right>$ \\
\noalign{\smallskip}
$\left| [211],\frac{1}{2},-\frac{1}{2},\frac{1}{3} \right>_{\eta}$ &=& 
$\left| \frac{1}{\sqrt{6}} (rgb-grb+brg-rbg+gbr-bgr)g \right>$ \\
\noalign{\smallskip}
$\left| [211],0,0,-\frac{2}{3} \right>_{\eta}$ &=& 
$\left| \frac{1}{\sqrt{6}} (rgb-grb+brg-rbg+gbr-bgr)b \right>$ \\
\noalign{\smallskip}
\end{tabular}
\end{ruledtabular}
\end{table*}

\bibliographystyle{apsrev4-1}
\bibliography{bibliography.bib}
\end{document}